\documentclass[]{interact}
\usepackage[caption=false]{subfig}
\usepackage[numbers,sort&compress,merge]{natbib}
\bibpunct[, ]{[}{]}{,}{n}{,}{,}
\renewcommand\bibfont{\fontsize{10}{12}\selectfont}
\usepackage{multirow,tabularx}
\usepackage{physics}

\begin{document}

\title{Applications of a perturbation-aware local correlation method to coupled
cluster linear response properties}
\author{\name{Ruhee D'Cunha\textsuperscript{a} and T. Daniel
Crawford\textsuperscript{a,b}\thanks{CONTACT T. Daniel Crawford. Email:
crawdad@vt.edu}}
\affil{\textsuperscript{a}Department of Chemistry, Virginia Tech, Blacksburg, VA
24061, USA; \textsuperscript{b}Molecular Sciences Software Institute, 1880 Pratt
Drive, Suite 1100, Blacksburg, VA 24060, USA} }

\maketitle

\begin{abstract}
We have investigated the efficacy of two recently proposed variations of the
pair-natural-orbital approach to reducing the scaling of coupled cluster property calculations.
In particular, we have extended our implementations of the PNO++ and combined PNO++ methods,
which make use of field-aware pair-densities to define the virtual-orbital spaces used to
describe electron correlation effects, in order to test their accuracy, efficiency, and
robustness on larger molecular systems than previously investigated.  For fluoroalkane chains
up to 1-fluoroheptane we find that the PNO++ and combined PNO++ methods yield smaller
localization errors in response properties than PNO for similarly compact virtual spaces, and,
while the PNO method performs better than the PNO++ method for correlation energies, the
combined PNO++ method recovers similar accuracy for correlation energies to the PNO method.
For more three-dimensional molecular structures such as $\alpha$- and $\beta-$pinenes, the PNO,
PNO++, and combined PNO++ methods all yield similar errors for response properties, whereas for
(\textit{S})-1-phenylethanol, the PNO method performs slightly better than the other two
approaches.  We also investigate the use of a product density to define the virtual space, as
well as two candidates for defining weak-pair contributions.
\end{abstract}


\section{Introduction}

Coupled-cluster theory is one of the most popular methods used for electronic
structure calculations, due to its expected accuracy and systematic
improvability\cite{Crawfordb, Bartlett2007}, which arise though its exponential
form of the wave function and convergence to the exact, nonrelativistic solution
to the electronic Schr\"odinger equation within a finite basis set.  However,
this performance comes at a substantial price, as coupled cluster suffers from
steep scaling of its computational cost with system size.  For example, coupled
cluster wave functions truncated at the single- and double-excitation (CCSD)
level scales as ${\cal O}(N)^6$, while including triples excitations (CCSDT)
increases the scaling to ${\cal O}(N)^8$, where $N$ is a measure of the size of
the system.\cite{Helgaker2008}

The high-degree polynomial scaling problem has been tackled by the development
of numerous methods over the last several decades, including fragmentation
approaches,\cite{Gordon2012, Epifanovsky2013, Li2004, Li2009} tensor
decompositions,\cite{Kinoshita2003, Koch2003, Hohenstein2012, Schutski2017,
Parrish2019, Pawowski2019}, localization schemes\cite{Pulay1983, Hampel1996,
Neese2009, Schutz2001b, Yang2012} and more, each with its own advantages and
disadvantages.  The main motivation for localization methods, in particular, is
that the dynamic electron correlation effects that coupled cluster aims to
capture are relatively short-ranged, and thus a localized basis would better
represent the sparsity present in the wave function than the usual delocalized
molecular orbital (MO) basis.  Local-correlation schemes to reduce the size of
the wave function parameter space, such as projected atomic orbitals
(PAOs)\cite{Pulay1983, Saebo1993, Hampel1996}, pair natural orbitals
(PNOs)\cite{Meyer1973,
Edmiston1966,Edmiston1968,Ahlrichs1975,Neese2009,Neese2009a}, and orbital
specific virtuals (OSVs)\cite{Yang2012} have been successfully applied to
coupled cluster ground state energies, and several methods have been developed
in order to apply them to excited state energies\cite{Crawford2002, Korona03,
Helmich2011, Dutta2016, Peng2018a, Frank2018a, Myhre2016}. However, their
application to response properties has been limited\cite{Gauss2000, Korona04,
Kats2007, Ledermuller2013, Schutz2015} due to the observed sensitivity of such
properties --- especially mixed electric-field/magnetic-field responses --- to
the definition of the correlation space\cite{Crawford2019}.

Russ and Crawford investigated an alternative domain-selection scheme for PAOs
that incorporated the perturbation into the virtual-orbital domain
definition\cite{Russ2004, Russ2008} and found that the computational crossover
points between canonical and local schemes lay at much larger molecules than for
ground-state energies.  Similarly, McAlexander and Crawford\cite{Mcalexander},
in a first application of PNOs to coupled-cluster linear-response properties,
found that, due to the reduced sparsity of the field-perturbed wave function in
the localized orbital basis, the sensitivity of response properties to
truncation of the wave function basis was much higher than for properties
requiring only the unperturbed state.  Kumar and Crawford\cite{Kumar2017},
exploring the effect of truncation of a natural orbital space on linear response
properties, saw an inverse relationship between the diffuseness of the orbitals
and their occupation number, and thus the elimination of low occupation number
orbitals in such schemes leads to the removal of more diffuse orbitals, which
are essential for accurate modeling of response properties.  Consideration of
the character of target electronically excited states led to the creation of
transition-specific natural orbitals by Høyvik, Myhre and Koch\cite{Hoyvik2017},
and the use of a similar approach by Baudin and Kristensen, in combination with
the local framework to reduce the scaling of CC2-level excitation energy
calculations without significant loss of accuracy\cite{Baudin2016, Baudin2017}.
H\"ofener and Klopper formed effective natural transition orbitals using ground
and excited state densities\cite{Hofener2017}, while Mester, Nagy, and Kállay
combined MP2 and CIS(D) densities to produce state-specific natural orbitals for
excitation energy calculations\cite{Mester2017, Mester2018}.

In our group's most recent work, the incorporation of external perturbations into the
construction of the virtual-orbital space for the PNO method led us to the creation and
exploration of the PNO++ approach for coupled cluster response properties.\cite{Crawford2019a,
DCunha2021} The PNO++ method described in Ref \citenum{DCunha2021} has shown encouraging results
for small, localizable molecular systems. In addition, the ``combined PNO++'' method, which
includes a fixed number of PNOs in the final orbital space, yields smaller localization errors in
the correlation energy than the PNO++ method alone --- as well as smaller errors in the response
properties studied than the PNO method alone. In our pilot, Python-based implementation of these
new methods, we used a canonical-MO formulation of the CC response equations, but then simulated
the effect of the local truncations by filtering out non-local contributions in each iterative
step.  (We have described such simulations in detail in previous
work.\cite{DCunha2021,Mcalexander}) However, in order for either method to be implemented at
production-level, validation is necessary on larger systems than are feasible in our original 
Python-based code.

The sensitive nature of response properties means that a larger proportion of the orbital space
needs to be kept to maintain accuracy in the value of the property.  However, a key efficiency
characteristic of the PNO method is aggressive truncation of the virtual space in order to avoid
increasing the computational cost of the initial integral transformation into the PNO basis.
Thus, a production-level code requires careful consideration as to an optimal method, including
the ability to truncate aggressively for larger, more delocalized systems.  The simulation code
described above, on the other hand, is essentially identical in cost to a canonical coupled
cluster calculation, because it only incorporates a single extra transformation (to filter out
non-local contributions) at each iteration.  A simulation code implemented in an optimized,
canonical-MO coupled cluster linear response code, such as the one present in
Psi4,\cite{Smith2020} provides an effective testing environment for further approximations
as well as improvements to the method for specific properties, as described below.

In this work, we apply the PNO++ and combined PNO++ methods as implemented as a simulation code
in Psi4 to larger organic molecules in order to validate the methods and compare them to the
performance of the PNO method for the same systems.  We also explore refinements to the PNO++
method, including a new formulation of the PNO++ density specifically for optical rotation
calculations, and the application of energy- and perturbation-based weak pair approximations to
reduce the computational expense of the methods.

\section{Theory and Computational Details}

\subsection{Larger Benchmark Calculations}

\begin{figure*}[h]
    \subfloat[]{
        \includegraphics[width=0.4\textwidth]{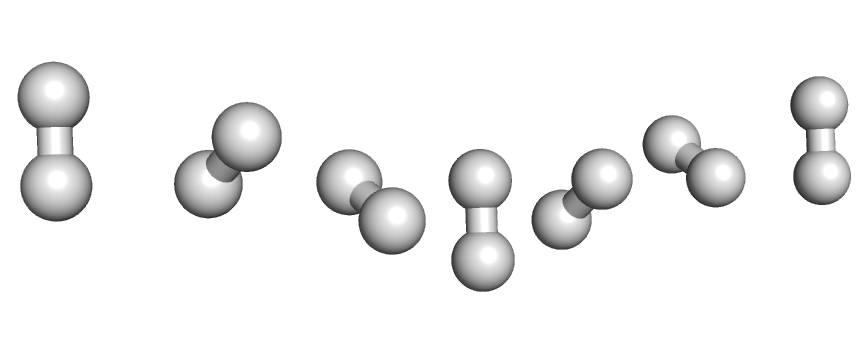}
    }
    \subfloat[]{
        \includegraphics[width=0.175\textwidth]{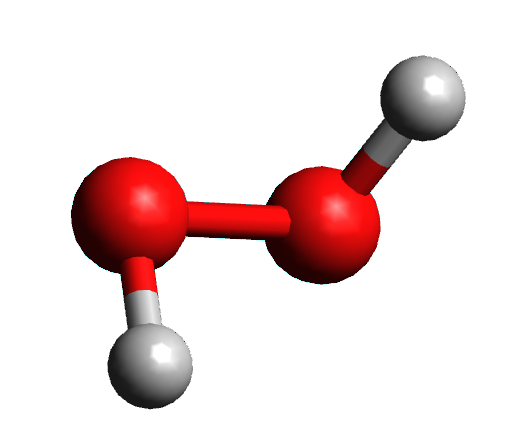}
    }
    \subfloat[]{
        \includegraphics[width=0.3\textwidth]{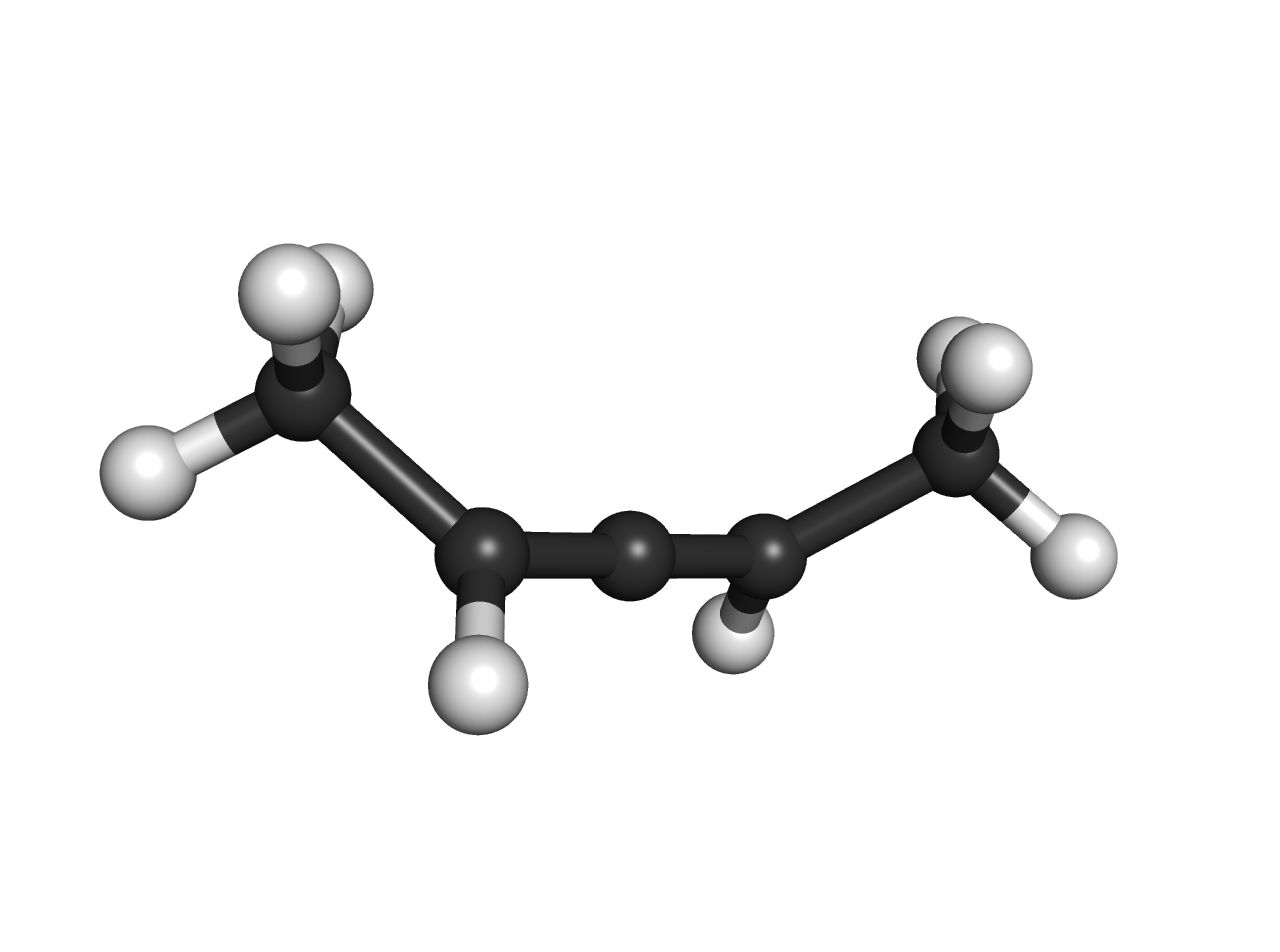}
    }
 \\
    \subfloat[]{
        \includegraphics[width=0.32\textwidth]{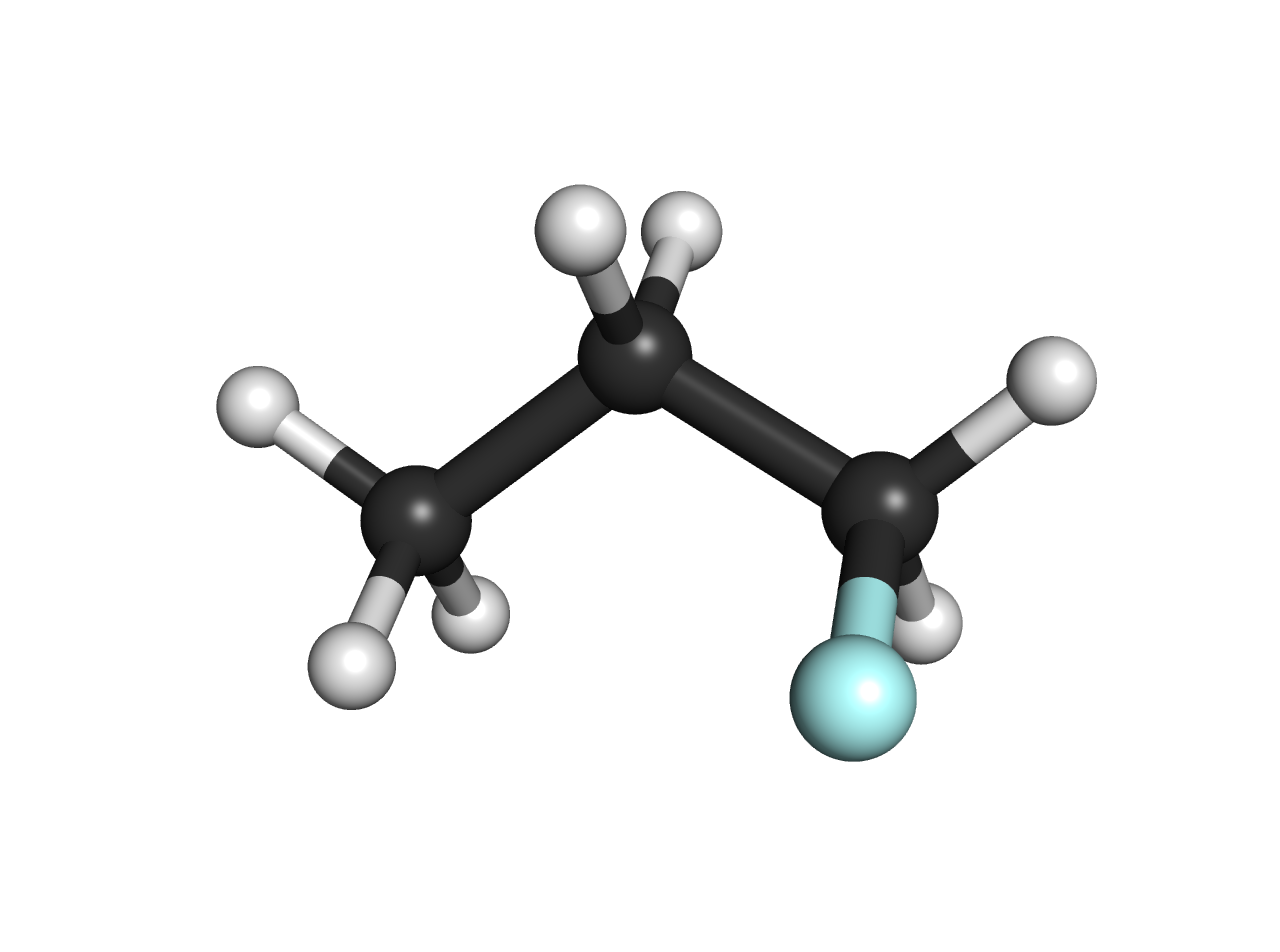}
    }
    \subfloat[]{
        \includegraphics[width=0.32\textwidth]{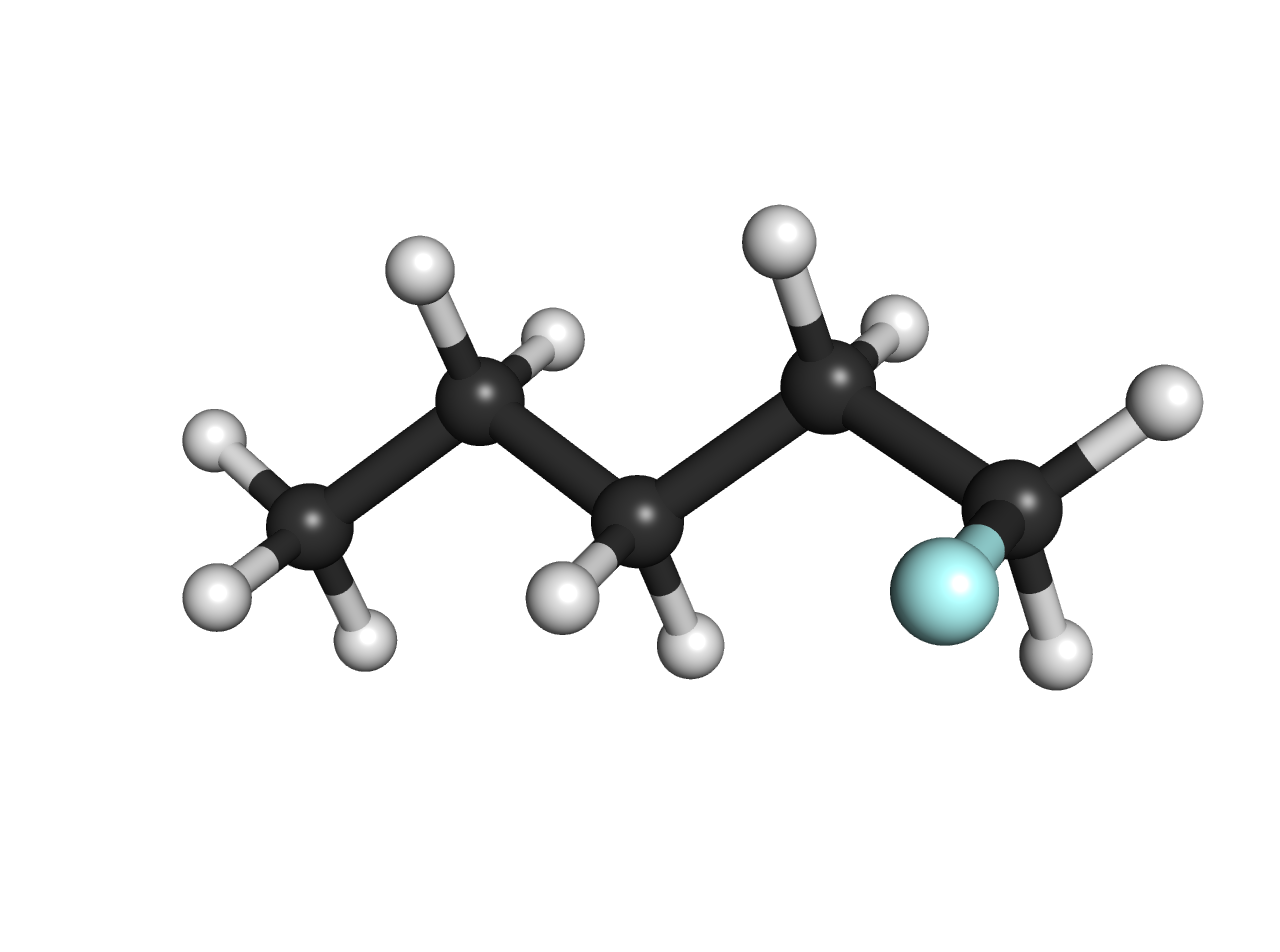}
    }
    \subfloat[]{
        \includegraphics[width=0.32\textwidth]{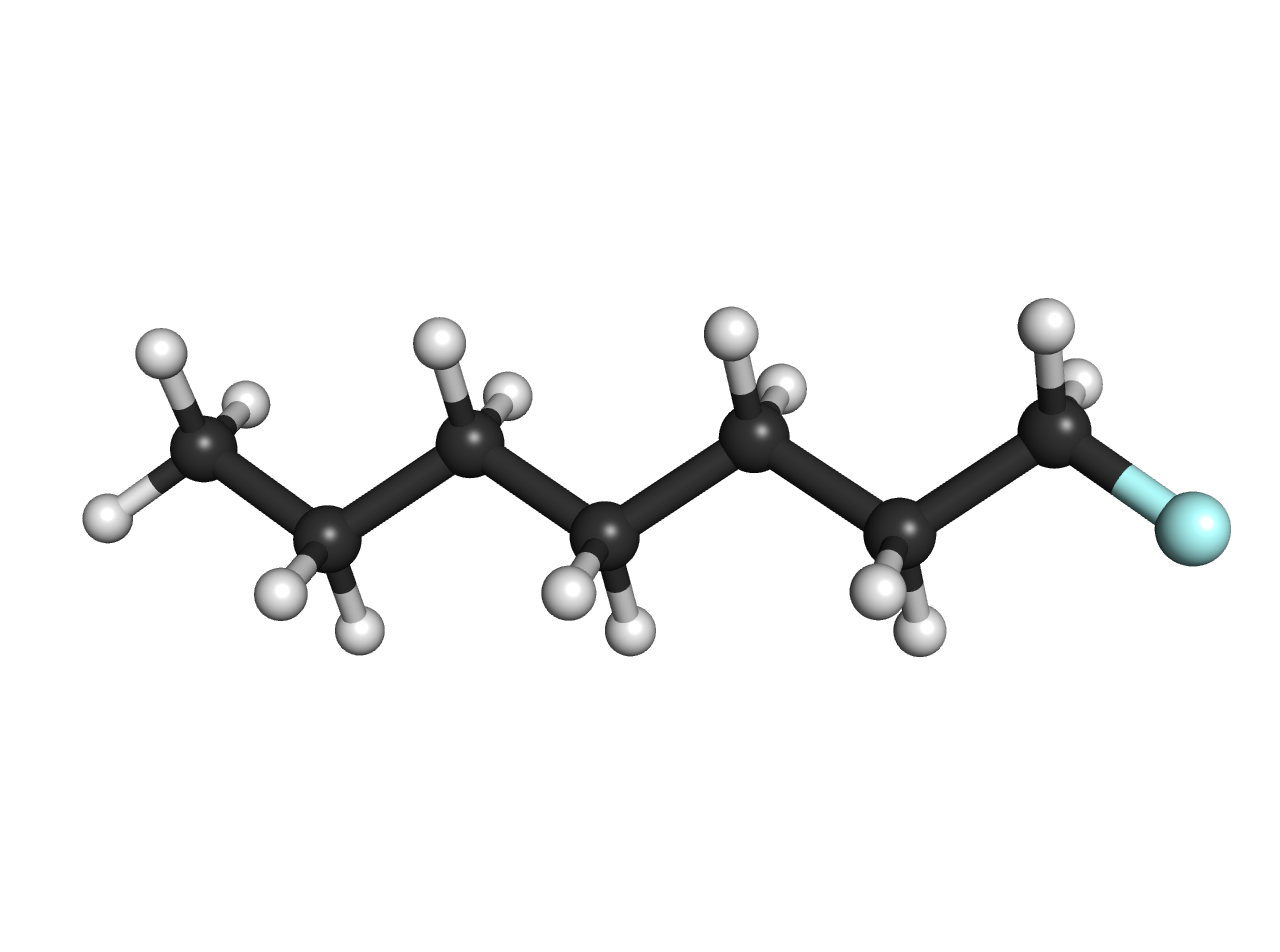}
    }
 \\
    \subfloat[]{
        \includegraphics[width=0.32\textwidth]{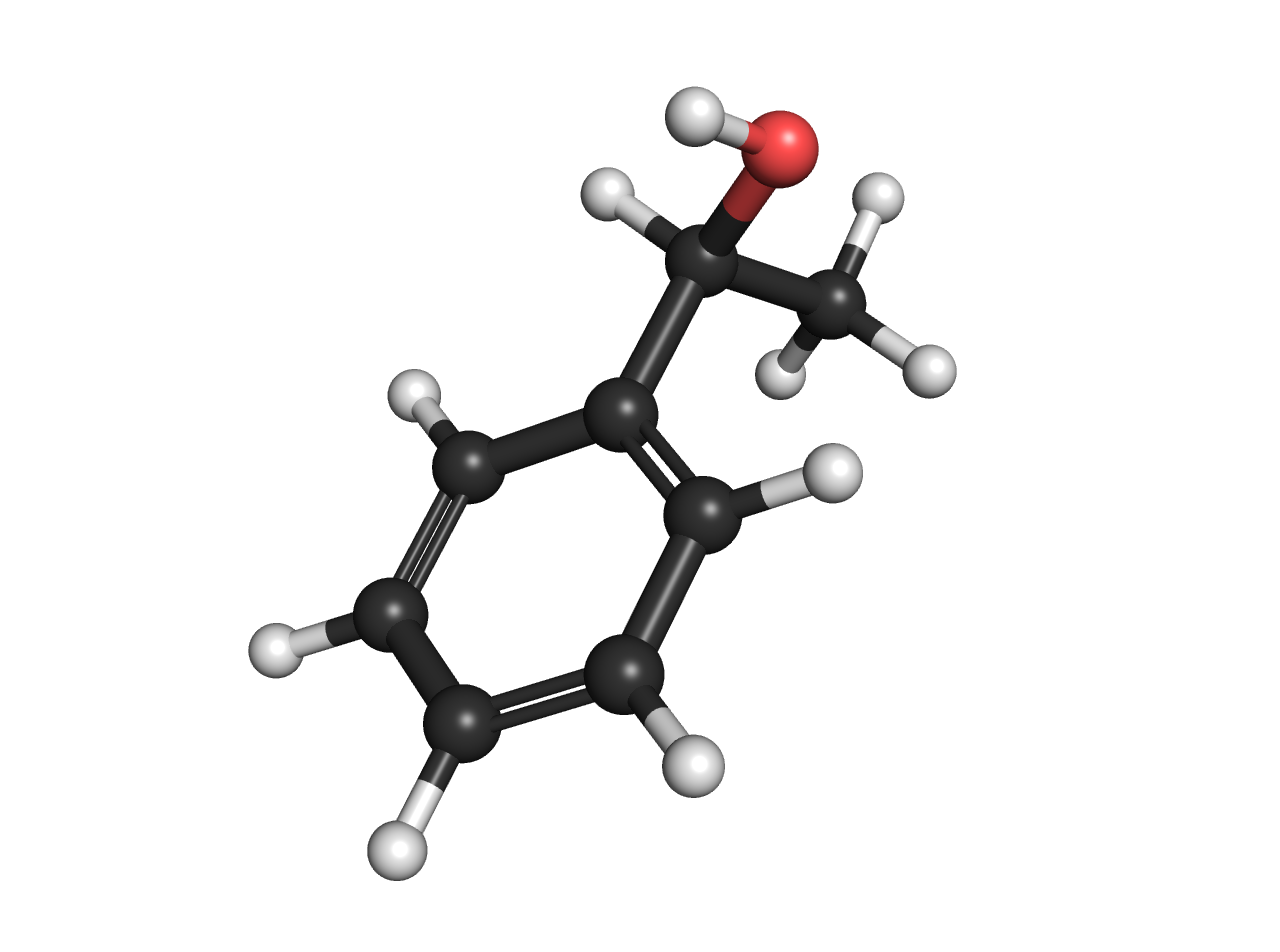}
    }
    \subfloat[]{
        \includegraphics[width=0.32\textwidth]{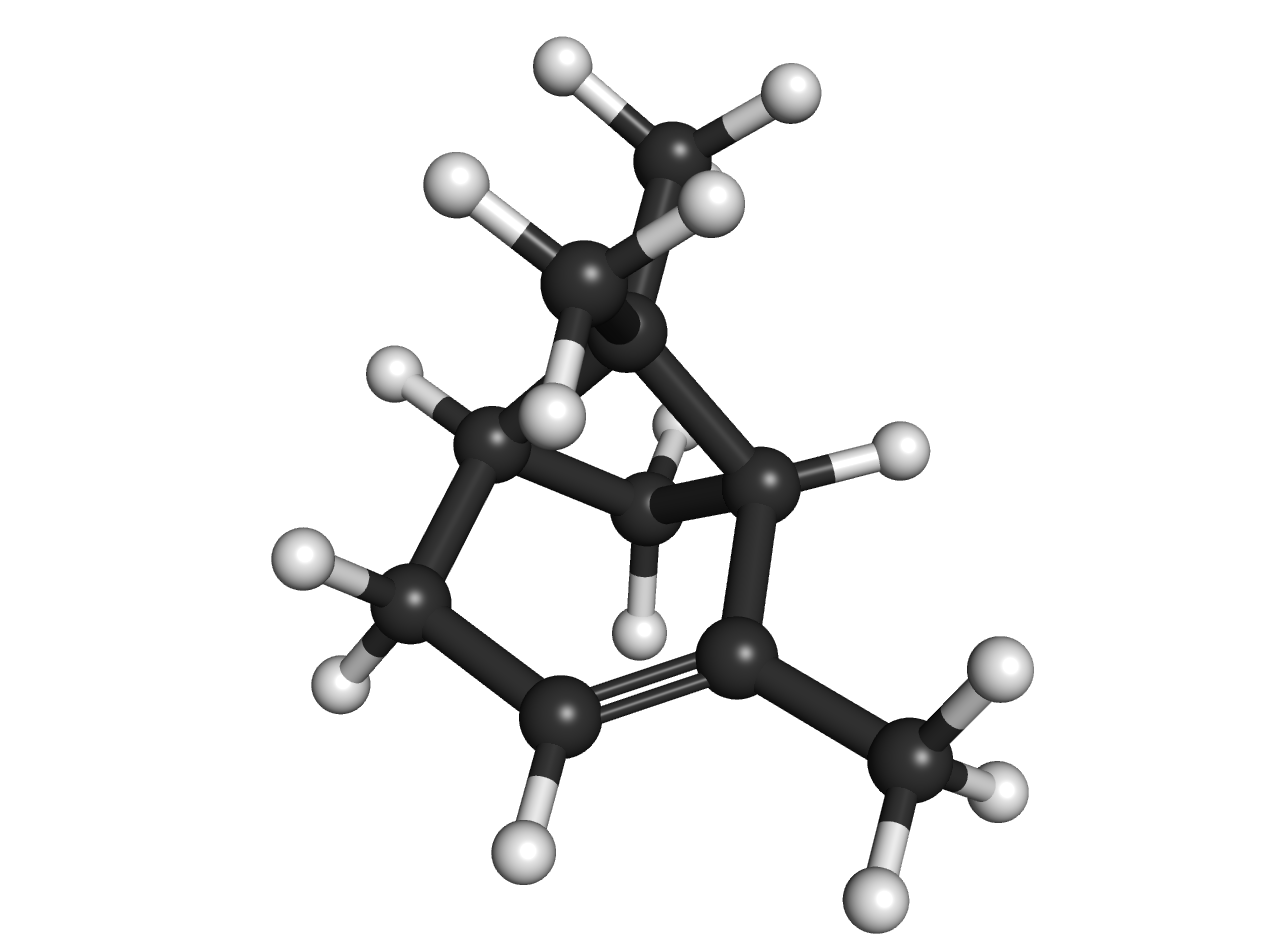}
    }
    \subfloat[]{
        \includegraphics[width=0.32\textwidth]{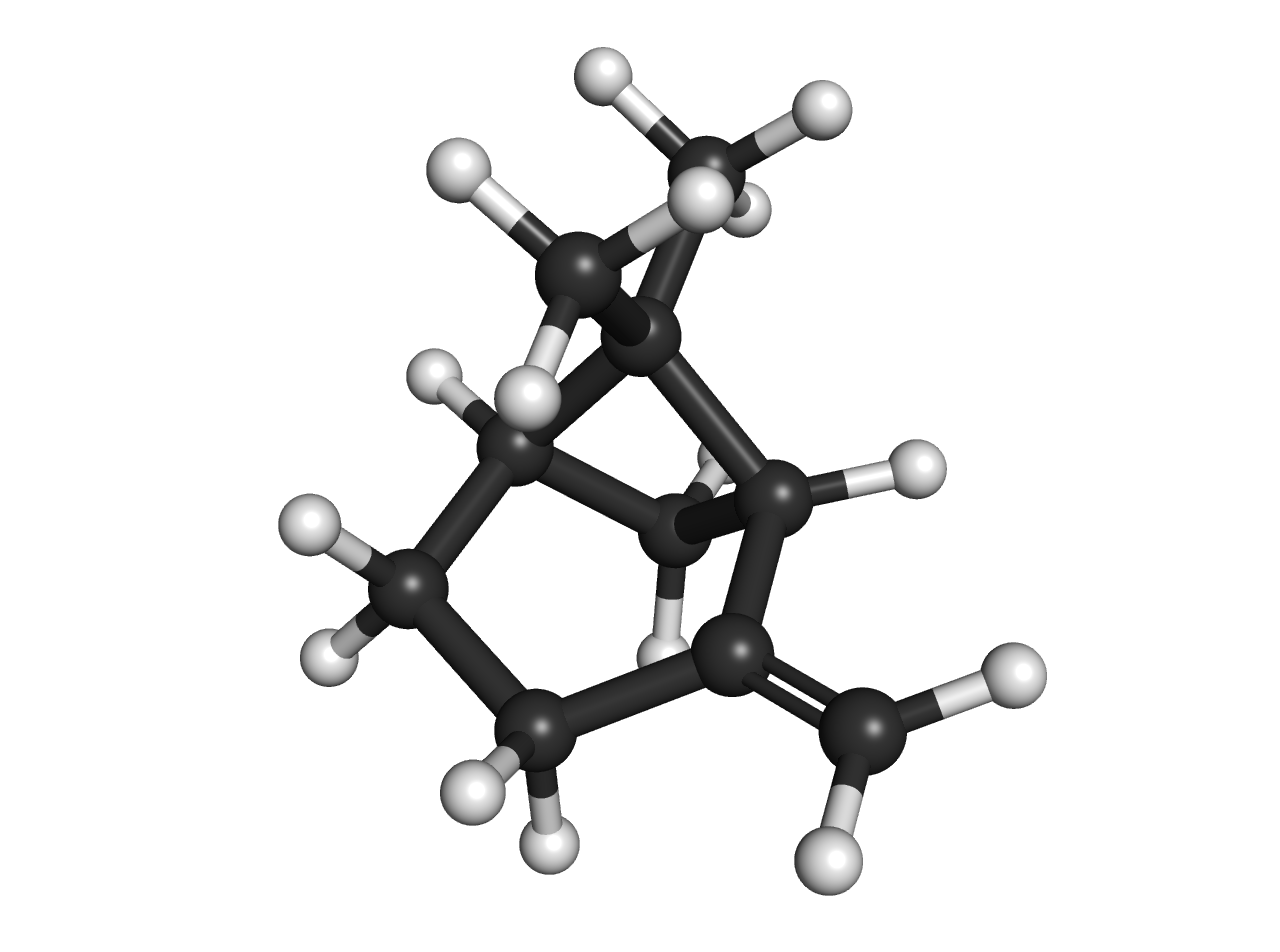}
    }
    \caption{Molecular systems studied in this work: (a) an (H$_2$)$_7$ helix, (b) H$_2$O$_2$, (c) $(P)$-1,3-dimethylallene (DMA) (d) (\textit{M})-1-fluoropropane, (e) (\textit{M})-1-fluoropentane, (f) (\textit{M})-1-fluoroheptane, (g) (\textit{S})-1-phenylethanol, (h) (1\textit{R},5\textit{R})-$\alpha$-pinene, and (i) (1\textit{R},5\textit{R})-$\beta$-pinene. All optimized geometries can be found in the Supporting Information.}
    \label{molecules}
\end{figure*}

Large-molecule benchmark calculations were carried out using an implementation of a local
correlation simulation code in a development version of the Psi4 program package.\cite{Smith2020}
In this code, non-local contributions to a given residual vector were filtered out at each
iteration in the coupled cluster amplitude, left-hand amplitude, and perturbed wave function
equations. Calculations on small molecule systems were carried out using a Python-based simulation
code. (For further details of the simulation see Ref \citenum{DCunha2021}.) The combined
PNO++ method had its $T_{cutPNO}$ threshold fixed to a reasonable value of $10^{-6}$.  All
molecular geometries were optimized at the B3LYP\cite{Becke1993a, Lee1988,
Stephens94:B3LYP}/aug-cc-pVDZ\cite{Dunning89,Kendall1992} level using Psi4's optking module. The
molecules studied in this work are given in Fig~\ref{molecules} and include two 
hydrogen molecule helices, H$_2$O$_2$, 1,3-dimethylallene (DMA), a set of
fluoroalkane chains, a phenyl-substituted alcohol, and two bicyclic alkenes.  Dipole
polarizabilities and specific rotations were computed using CCSD linear
response\cite{Pedersen1997,Koch1990b} and an external field wavelength of 589 nm, with and
without the local simulation applied.  The specific rotation was computed using both the length
and the velocity gauge representation of the dipole moment operator, with a shift to zero field
frequency to obtain the modified velocity gauge value of the specific
rotation.\cite{Pedersen1997a,Pedersen2004} All response calculations used the augmented
correlation-consistent double zeta basis set of Dunning and coworkers,
aug-cc-pVDZ.\cite{Woon1993}  

\subsection{Product-based densities}
\label{product_densities}

Pair densities in the PNO and PNO++ approaches are formed using MP2-level amplitudes in the form
of the MP2 reduced density matrix. Spaces of virtual natural orbitals are then formed by
diagonalization of the pair densities, with the PNO++ density including some measure of the
external perturbation in order to improve the compactness of the space for response properties.
However, the PNO++ density, written as
\begin{equation}
    \bm{D}^{ij}(B,\omega) = \frac{2}{1+\delta_{ij}} \qty(\bm{X}^{ij}_B \bm{\widetilde{X}}^{ij\dagger}_    B + \bm{X}^{ij\dagger}_B \bm{\widetilde{X}}^{ij}_B)
\end{equation}
only contains dependence on a single perturbation $B$, e.g. the electric dipole operator
$\hat{\mu}$ or the magnetic dipole operator $\hat{m}$.  The optical rotation tensor
$\beta(\omega)$, on the other hand, requires both of these operators:
\begin{equation}
    \beta(\omega) = \textrm{Im} \langle \langle \hat{\mu}; \hat{m} \rangle \rangle_{\omega}.
    \label{mixed}
\end{equation}

The linear response function can further be formulated in terms of the first-order perturbed
density as
\begin{equation}
    \langle \langle \hat{\mu} ; \hat{m} \rangle\rangle_{\omega} = \frac{1}{2} \hat{P}(\hat{\mu},
    \hat{m})\left[\sum_{p q} {\hat{\mu}}_{p q}\left[D_{p
    q}^{\hat{m}^{\omega}}\right]^{(1)}\right],
\end{equation}
where the permutation operator $\hat{P}$ ensures that the linear response function contains both
$[D_{p q}^{\hat{\mu}^{\omega}}]^{(1)} $ and $[D_{p q}^{\hat{m}^{\omega}}]^{(1)}$.  Thus, a
reasonable form of the pair density used for creating the space could include an element-wise
(Hadamard) product of the perturbed pair densities containing $\hat{\mu}$ and $\hat{m}$, i.e.,
\begin{equation}
    \bm{D}^{ij} = \bm{D}^{ij} (\hat{\mu},\omega) \circ \bm{D}^{ij}  (\hat{m},\omega).
\end{equation}
Virtual orbitals created using this product density should, in principle, contain more
information about the response of the system in the case of the mixed response function in
Eq~(\ref{mixed}) above, leading to a more compact space for the specific rotation.


\subsection{Weak Pairs}

The sparsity of the localized occupied-MO basis can be exploited by treating at a lower level or
neglecting certain sets of PNOs. For example, those PNOs that correspond to the pairs of occupied
orbitals $ij$ whose approximated contributions to the quantity of interest are smaller than a
predetermined cutoff can be neglected. These contributions are used as a measure of the
importance of those pairs, and neglected pairs are known as ``weak pairs.'' For the PNO method, this
criterion takes the form
\begin{equation}
    |\epsilon_{ij}| < T_{cutPairs},
\end{equation}
where $\epsilon_{ij}$ is defined as the pair correlation energy as typically computed at the MP2
level of theory. The method works by neglecting or otherwise treating with a less expensive
method the pairs whose integrals,
\begin{equation}
    \int \phi_i^*(r_1) \phi_j^*(r_2) \frac{1}{r_{12}} \phi_a(r_1) \phi_b(r_2) dr_1 dr_2,
\end{equation}
are negligible, and thus whose contribution to the energy is also expected to be negligible.
Here, the pairs $ij$ are occupied, with the pairs $ab$ being virtual orbitals. This is true when
the orbitals are sufficiently distant spatially to have minimal overlap. This works well for
localized occupied orbitals in large molecules, and thus the number of pairs and therefore the
total number of wave function parameters to be computed and stored can be reduced to a much more
computationally efficient level.

Previous studies neglecting weak pairs using the energy criterion found that adding a second
truncation threshold in addition to the virtual-space truncation led to larger errors in response
properties at similar fractions of the space kept.  A similar perturbation-including criterion
can be formulated as
\begin{equation}
    |\bar{\mu}_{ij}| <T_{cutPairs},
\end{equation}
where
\begin{equation}
    \bar{\mu}_{ij} = \sum_{ab} \bar{\mu}_{ij}^{ab},
\end{equation}
with $\bar{\mu}_{ij}^{ab}$ being the similarity-transformed perturbation using MP2-level doubles amplitudes,
\begin{equation}
    \bar{\mu}_{ij} = e^{-T_2} \mu e^{T_2}.
\end{equation}
In this work, we neglect contributions from the weak pairs, determined either using the MP2 pair
correlation energy or the MP2-level similarity transformed perturbation, in order to quantify the
loss in accuracy at various truncation thresholds.  The goal is to combine this
perturbation-including weak pair criterion with the PNO++ method in order to truncate the space
even further than in previous work and attempt to maintain accuracy in the correlation energy as
well as response properties.

\section{Results and Discussion}

\subsection{Larger Benchmark Calculations}

\subsubsection{Fluoroalkane Chains}

\begin{figure*}[htbp]
    \centering
    \subfloat[]{
        \includegraphics[width=0.32\textwidth]{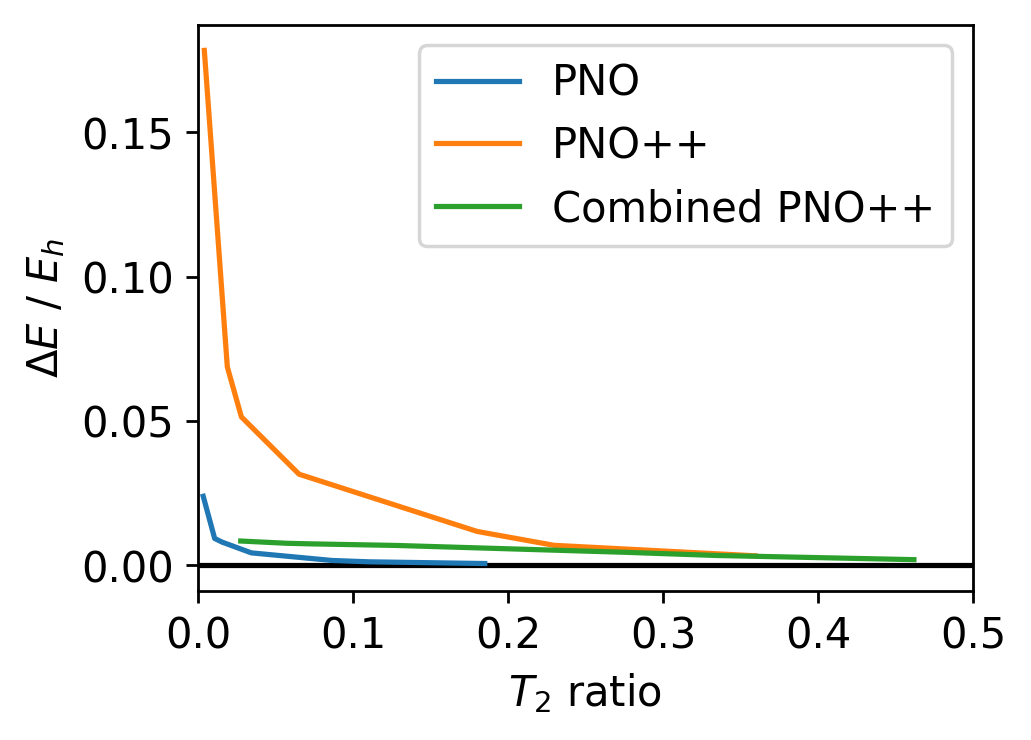}
    }
    \subfloat[]{
        \includegraphics[width=0.32\textwidth]{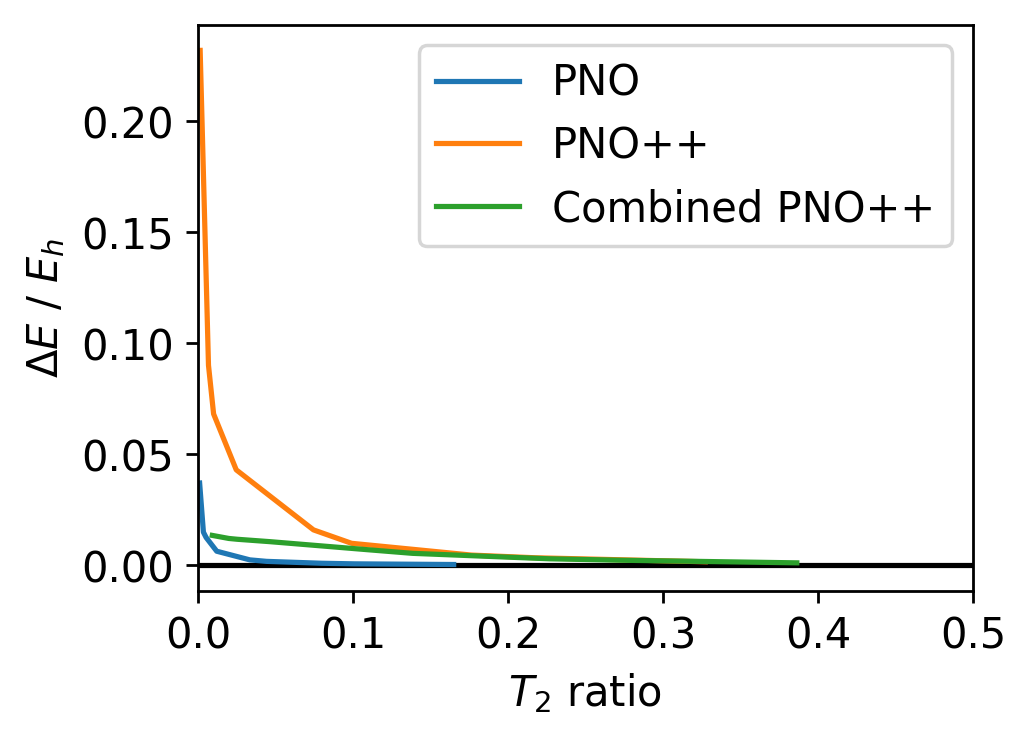}
    }
    \subfloat[]{
        \includegraphics[width=0.32\textwidth]{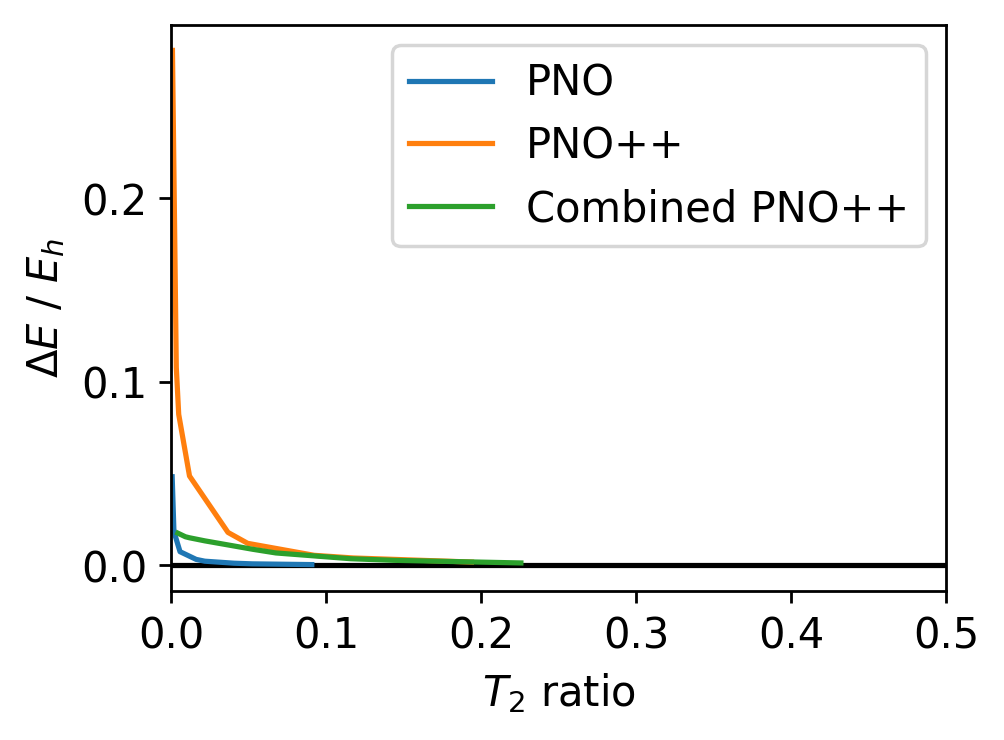}
    }
    \caption{Truncation errors in CCSD correlation energy in Hartree for (a) (\textit{M})-1-fluoropropane, (b) (\textit{M})-1-fluoropentane and (c) (\textit{M})-1-fluoroheptane systems, computed using the aug-cc-pVDZ basis set.}
    \label{falkane_en}
\end{figure*}

\begin{figure*}[htbp]
    \centering
    \subfloat[]{
        \includegraphics[width=0.32\textwidth]{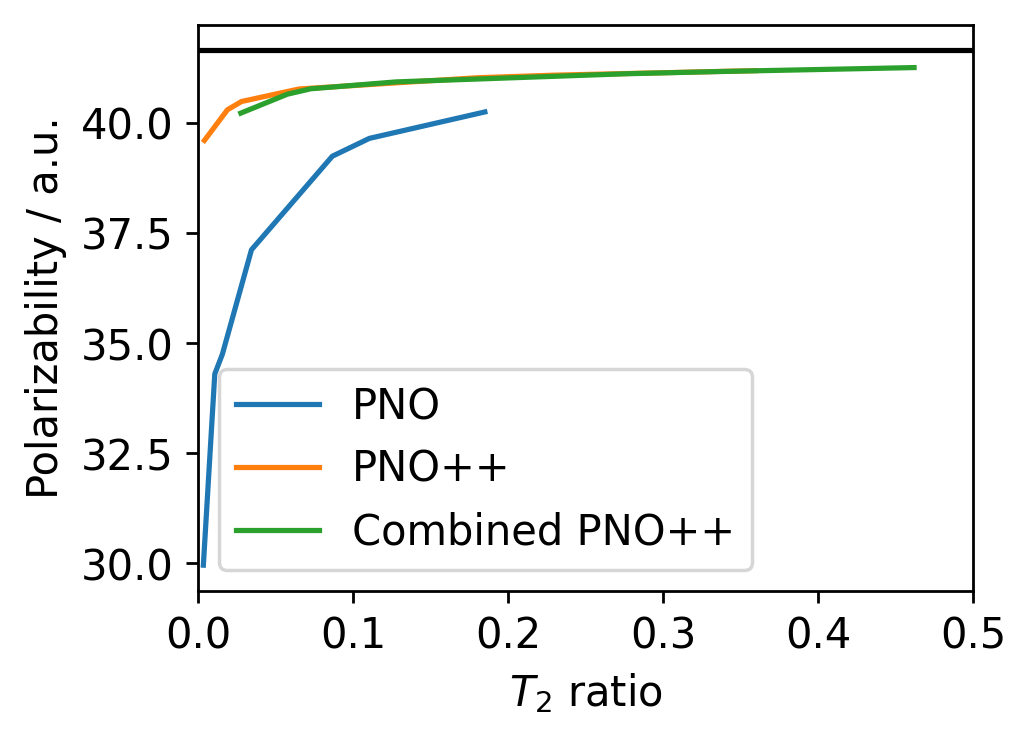}
    }
    \subfloat[]{
        \includegraphics[width=0.32\textwidth]{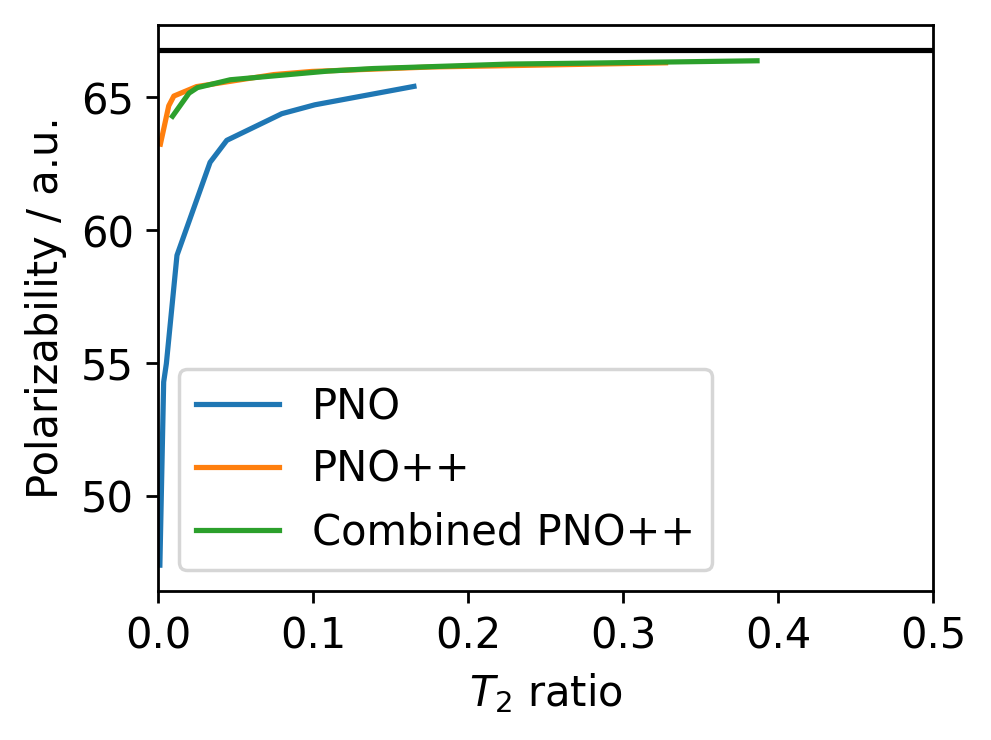}
    }
    \subfloat[]{
        \includegraphics[width=0.32\textwidth]{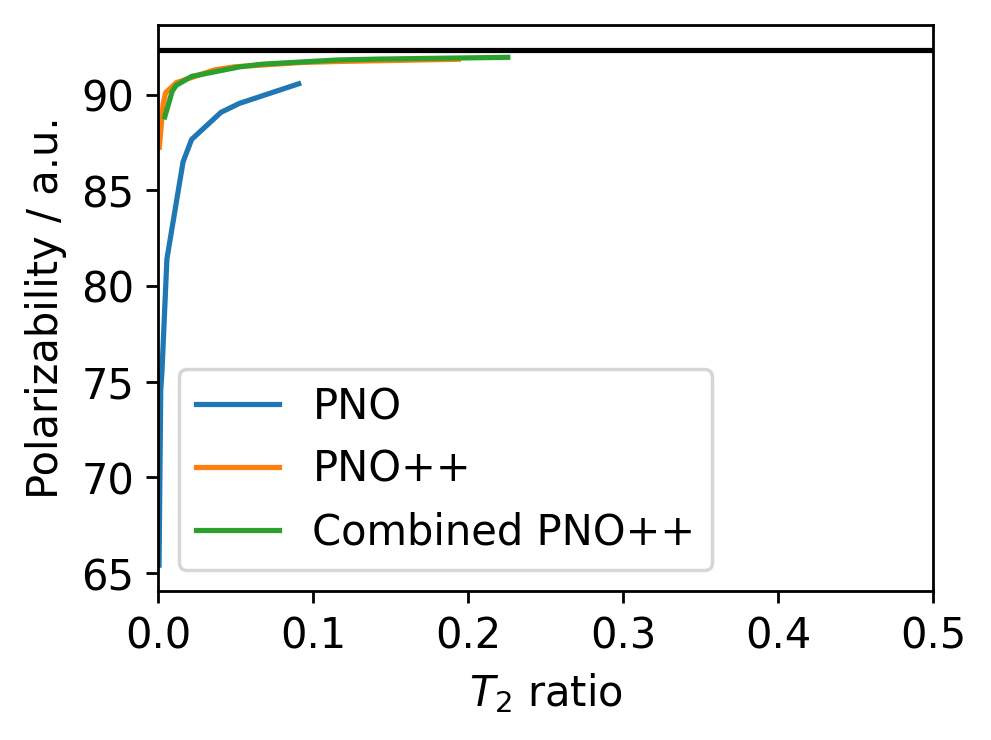}
    }
    \caption{CCSD linear response dynamic polarizabilities at 589 nm in a.u. for (a) (\textit{M})-1-fluoropropae, (b) (\textit{M})-1-fluoropentane and (c) (\textit{M})-1-fluoroheptane systems, computed using the aug-cc-pVDZ basis set.}
    \label{falkane_polar}
\end{figure*}

\begin{figure*}[htbp]
    \centering
    \subfloat[]{
        \includegraphics[width=0.32\textwidth]{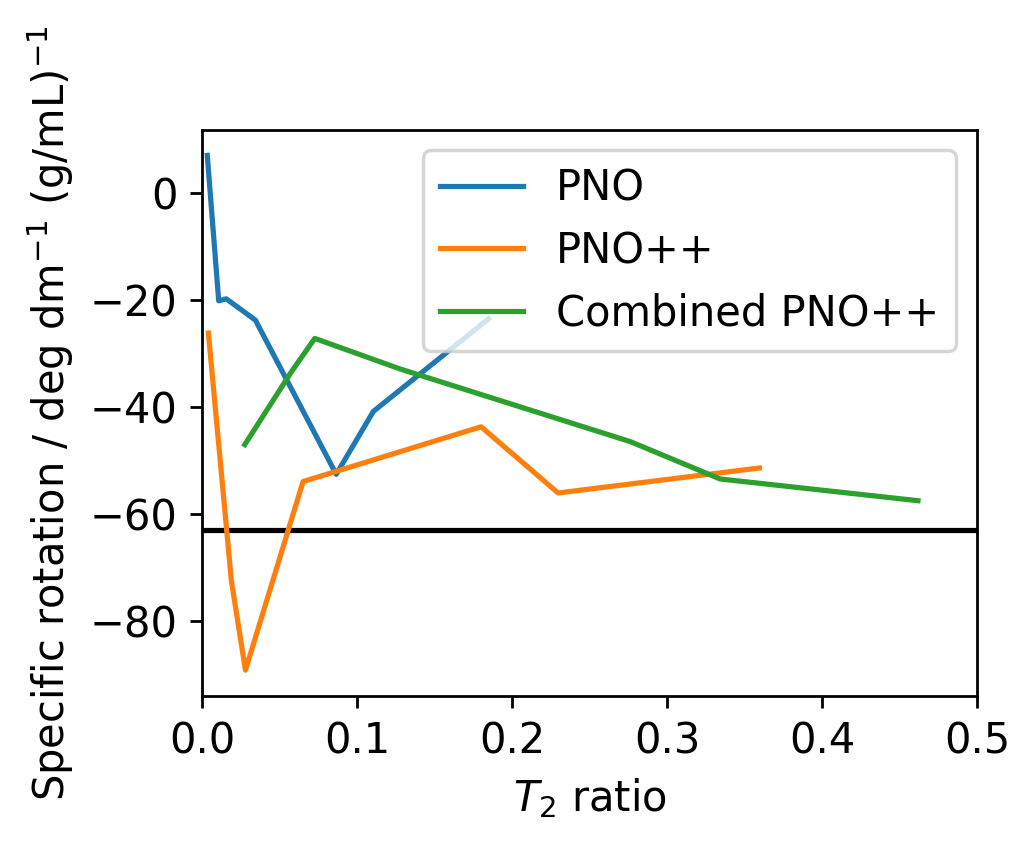}
    }
    \subfloat[]{
        \includegraphics[width=0.32\textwidth]{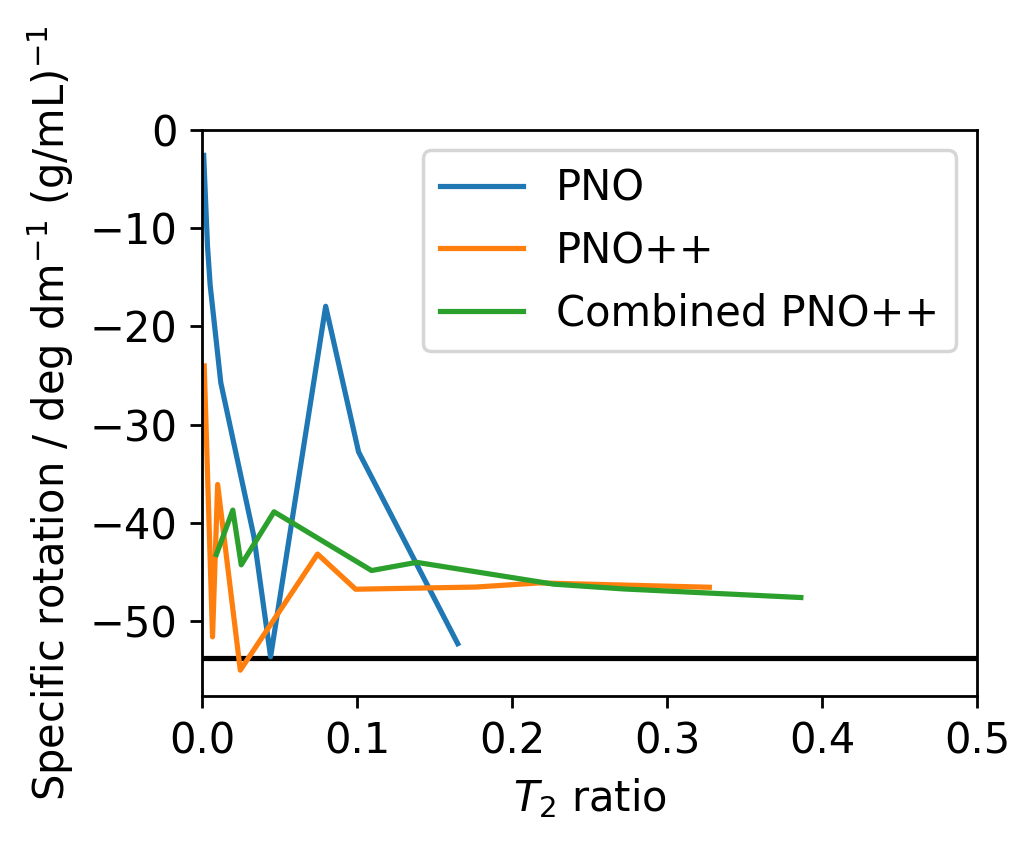}
    }
    \subfloat[]{
        \includegraphics[width=0.32\textwidth]{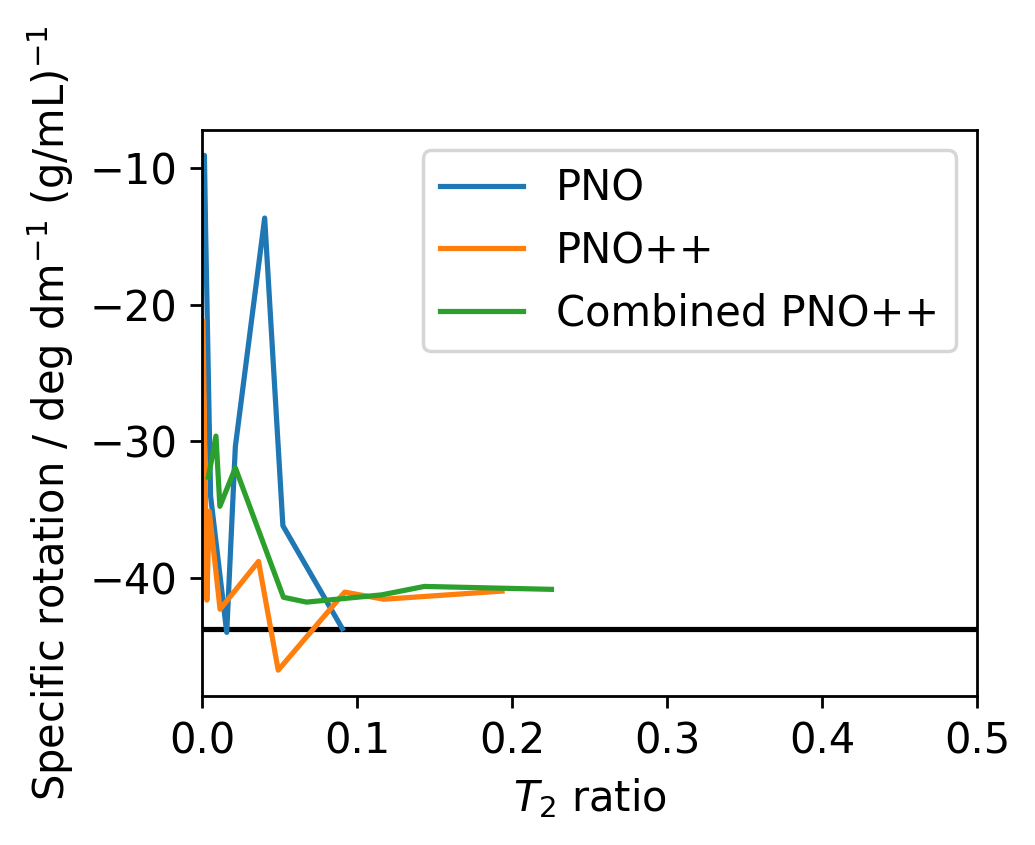}
    }
    \caption{CCSD specific rotations in deg dm$^{-1}$ (g/mL)$^{-1}$ at 589 nm for (a) (\textit{M})-1-fluoropropane, (b) (\textit{M})-1-fluoropentane and (c) (\textit{M})-1-fluoroheptane systems, computed using the aug-cc-pVDZ basis set.}
    \label{falkane_mvg}
\end{figure*}

Figures \ref{falkane_en}-\ref{falkane_mvg} illustrate the effect of truncation of the PNO, PNO++
and combined spaces on the errors in correlation energy, dipole polarizabilities and specific
rotations, respectively, for 1-fluoropropane, -pentane and -heptane. The systems are linear and
can provide information about the effect of an increasing linear system size on the truncation
errors in energy and properties.  In each case, the localization errors are plotted as a function
of the $T_2$ ratio, which is the ratio of the number of truncated $T_2$ amplitudes to the full
untruncated number and thus can be used as a measure of the potential computational savings
expected by the corresponding truncation of the space. 

Across the substituted propane, pentane, and heptane molecular species, we see that the
truncation errors in correlation energy are larger for the PNO++ method, with the PNO method
achieving values within chemical accuracy of the CCSD reference at $T_2$ ratios of 0.11, 0.04 and
0.02 respectively, while the PNO++ requires $T_2$ ratios of more than 0.36, 0.22 and 0.19 for the
same result.  The combined method, however, with a set threshold for unperturbed PNOs to be
included in the space, regains a significant portion of the accuracy of the PNO method at all
truncations. 

Figure \ref{falkane_polar} depicts the convergence in the values of the dipole polarizabilities
with respect to the $T_2$ ratio, with the error for the PNO method being larger than that for the
PNO++ and combined methods at each truncation. The PNO++ and combined methods show a
slowly-tapering error, reaching within 5\% of the reference value at a $T_2$ ratio of 0.03 but
naturally requiring a larger amount of the space for tighter convergence to the reference value.

Specific rotations computed for the fluoroalkane systems, which are optically active due to a
twist in the methyl group containing the fluorine away from the molecular mirror plane, can be
seen in Figure \ref{falkane_mvg}.  The data exhibit better convergence for the PNO++ and combined
PNO++ methods as compared to the PNO method, at $T_2$ ratios below 0.2 for the smaller
1-fluoropropane and 1-fluoropentane systems, and below 0.1 for the larger 1-fluoroheptane system.
The large fluctuations in specific rotation values seen with PNOs become smaller fluctuations
with the PNO++ and combined PNO++ methods, an observation that aligns with the results of our
previous study\cite{DCunha2021}.  However, rotations computed at $T_2$ ratios below 0.5 are slow
to converge to the reference value for both the PNO++ and the combined PNO++ methods.  Rotations
computed using the length-gauge representation of the electric dipole moment operator show faster
convergence to the reference value, results which are specific to this set of molecules and are
not reproduced for any other system studied (cf.\ Supporting Information).
While we do not observe full convergence of the linear response properties to the reference CCSD
value is not seen at $T_2$ ratios below 0.5, the results indicate that the PNO++ and combined
PNO++ methods still perform significantly better in terms of reducing the cost of property
calculations than the PNO method for this set of linear systems.

\subsubsection{$\alpha$- and $\beta$-pinene}

\begin{figure*}[htbp]
    \centering
    \subfloat[]{
        \includegraphics[width=0.32\textwidth]{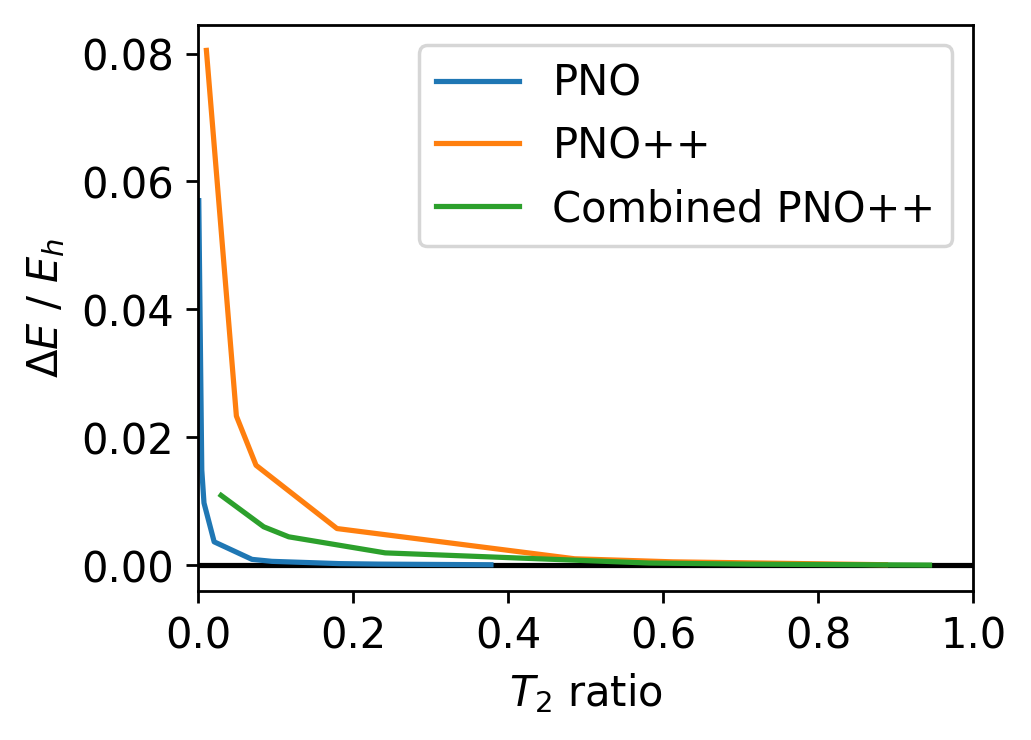}
    }
    \subfloat[]{
        \includegraphics[width=0.32\textwidth]{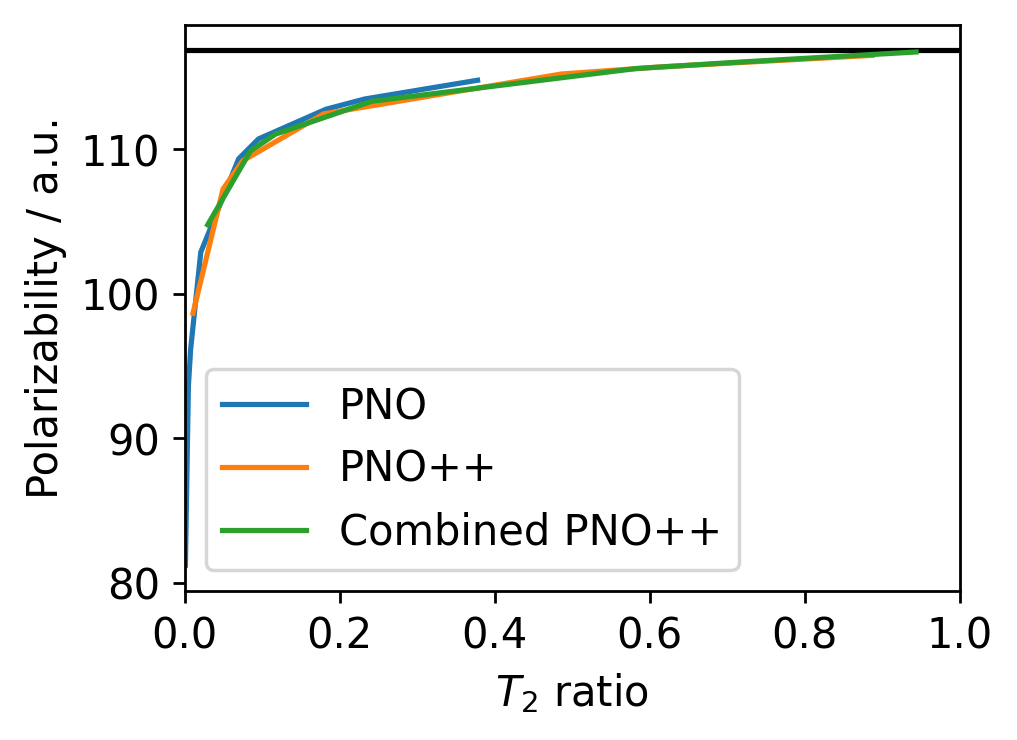}
    }
    \subfloat[]{
        \includegraphics[width=0.32\textwidth]{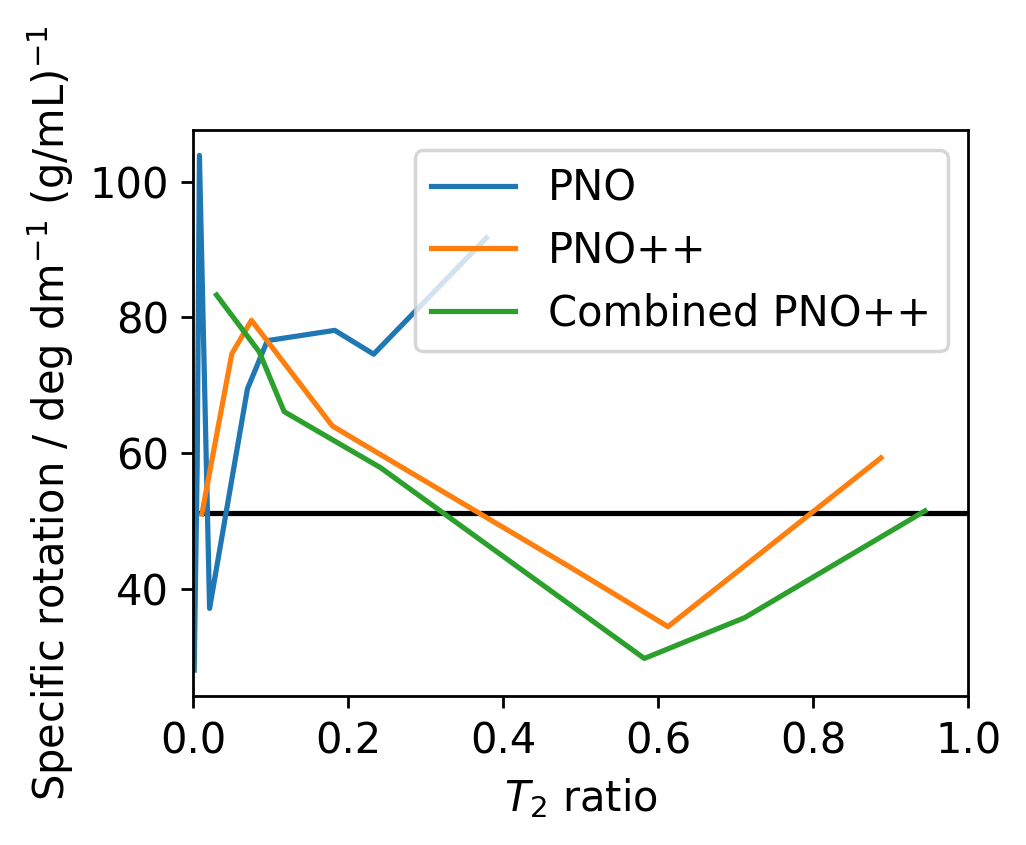}
    }
    \caption{ (a) Truncation errors in CCSD correlation energy in Hartree, (b) CCSD dynamic polarizabilities at 589 nm in a.u., and (c) CCSD specific rotations in deg dm$^{-1}$ (g/mL)$^{-1}$ at 589 nm for (1\textit{R},5\textit{R})-$\alpha$-pinene, computed using the aug-cc-pVDZ basis set.}
    \label{apinene}
\end{figure*}

\begin{figure*}[htbp]
    \centering
    \subfloat[]{
        \includegraphics[width=0.32\textwidth]{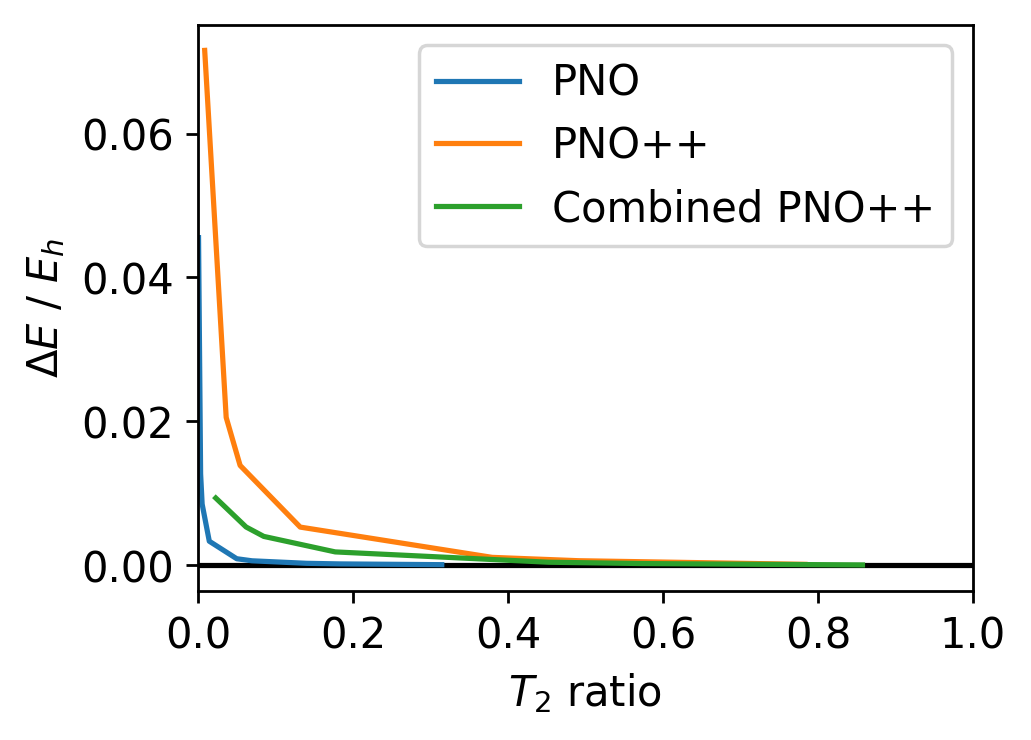}
    }
    \subfloat[]{
        \includegraphics[width=0.32\textwidth]{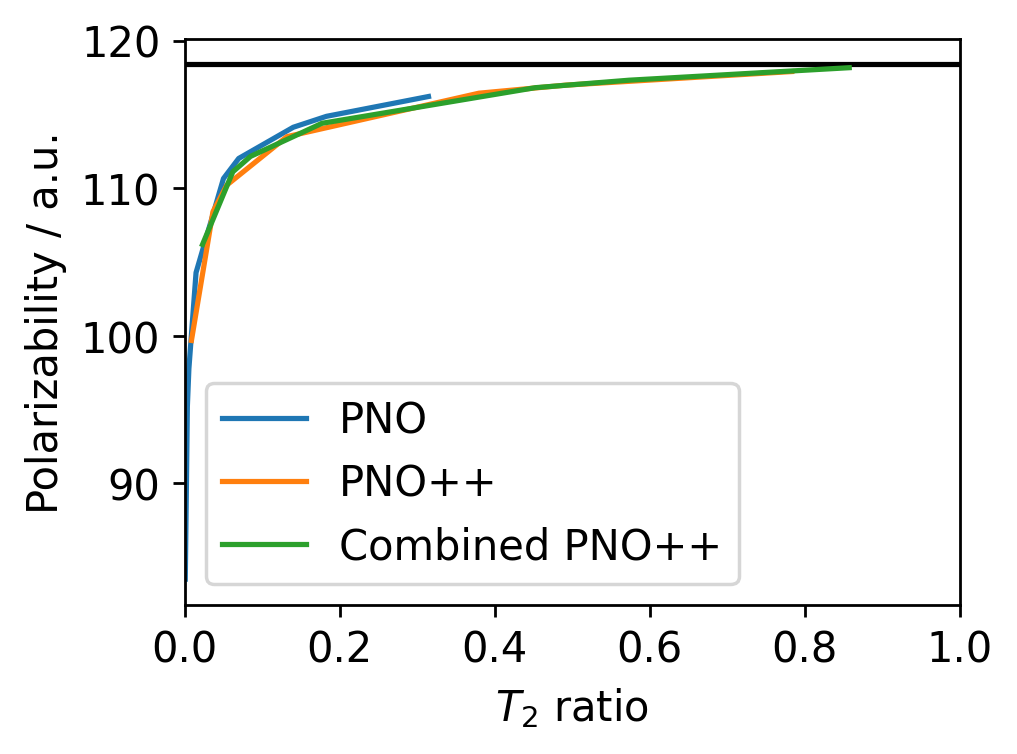}
    }
    \subfloat[]{
        \includegraphics[width=0.32\textwidth]{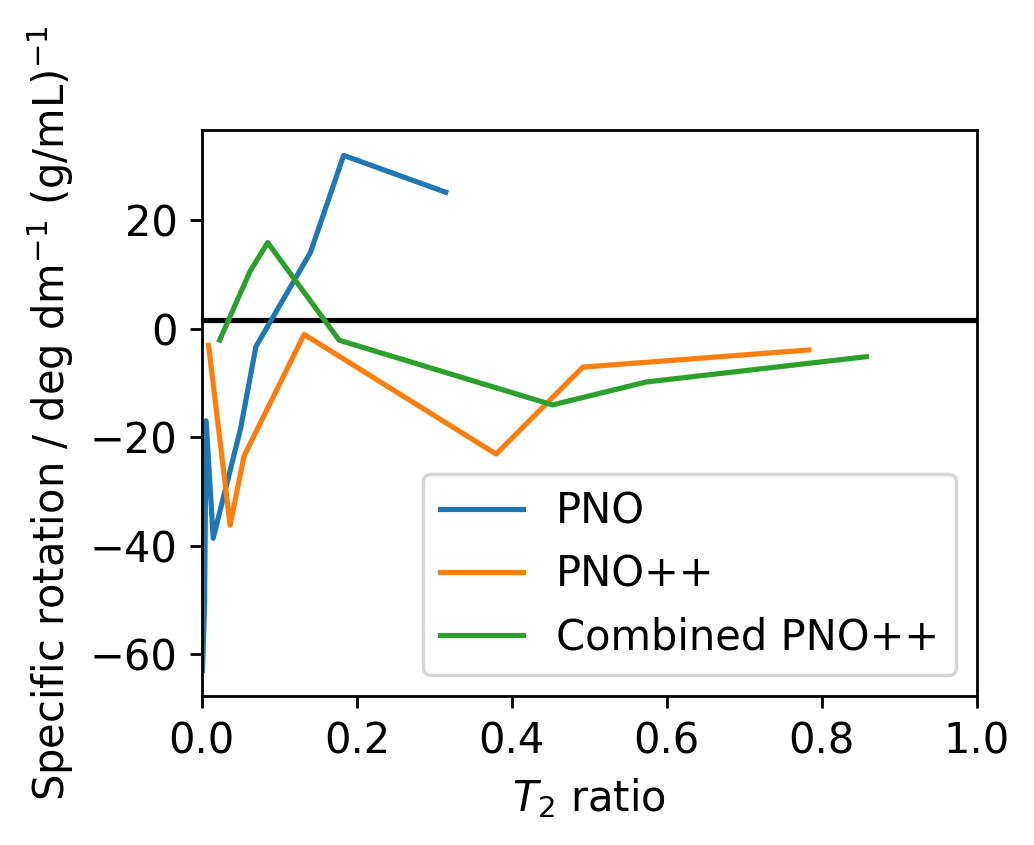}
    }
    \caption{ (a) Truncation errors in CCSD correlation energy in Hartree, (b) CCSD dynamic polarizabilities at 589 nm in a.u., and (c) CCSD specific rotations in deg dm$^{-1}$ (g/mL)$^{-1}$ at 589 nm for (1\textit{R},5\textit{R})-$\beta$-pinene, computed using the aug-cc-pVDZ basis set.}
    \label{bpinene}
\end{figure*}

Figures \ref{apinene} and \ref{bpinene} depict the effect of wave function truncation with the
PNO, PNO++ and combined PNO++ methods on correlation energies and linear response properties
computed for the systems of, respectively, (1\textit{R},5\textit{R})-$\alpha$-pinene and
(1\textit{R},5\textit{R})-$\beta$-pinene (Fig.\ \ref{molecules}).  Errors in correlation energies
[Fig.\ \ref{apinene}(a) and \ref{bpinene}(a)] follow the same trend as seen for the fluoroalkane
systems earlier, with the PNO method having much smaller errors in comparison with the PNO++ at
large truncations, and the combined method recovering similar accuracy at the same truncations.

Interestingly, dipole polarizabilities for both pinene systems [panel (b) in each figure]
calculated using the PNO, PNO++ and combined methods have similar errors at similar $T_2$ ratios.
This is in contrast to the results seen with the fluoroalkanes and indicates that all three
spaces show similar levels of compactness for the polarizability for these bicyclic molecules.

Figures \ref{apinene}(c) and \ref{bpinene}(c) contain specific rotations as a function of the
$T_2$ ratio. The pinene systems differ only by the placement of the double bond, but their
specific rotations calculated using CCSD linear response differ significantly, with the modified
velocity gauge value for $\alpha$-pinene being 51.19 deg dm$^{-1}$ (g/mL)$^{-1}$, and the same
for $\beta$-pinene being 1.72 deg dm$^{-1}$ (g/mL)$^{-1}$.  We see that for $\alpha$-pinene, all
three methods show oscillations around the reference value, with the error being fairly large up
to the $T_2$ ratio of 1.0, indicating that even small truncations of the space introduce a large
error (of up to 33\% at a $T_2$ ratio of 0.6) in the specific rotation value.  The small value of
the specific rotation of $\beta$-pinene introduces additional problems, as a negative truncation
error at small $T_2$ ratios means that the sign of the specific rotation for the PNO and PNO++
methods is incorrect at all $T_2$ ratios considered. We see this in Figure \ref{bpinene}(c). For
both pinene systems, while the error using the PNO method is larger than that using the PNO++
method at several $T_2$ ratios, neither the PNO++ nor the combined method can be used to reliably
maintain accuracy in the specific rotation value. 

\subsubsection{(\textit{S})-1-phenylethanol}

\begin{figure*}[htbp]
    \centering
    \subfloat[]{
        \includegraphics[width=0.32\textwidth]{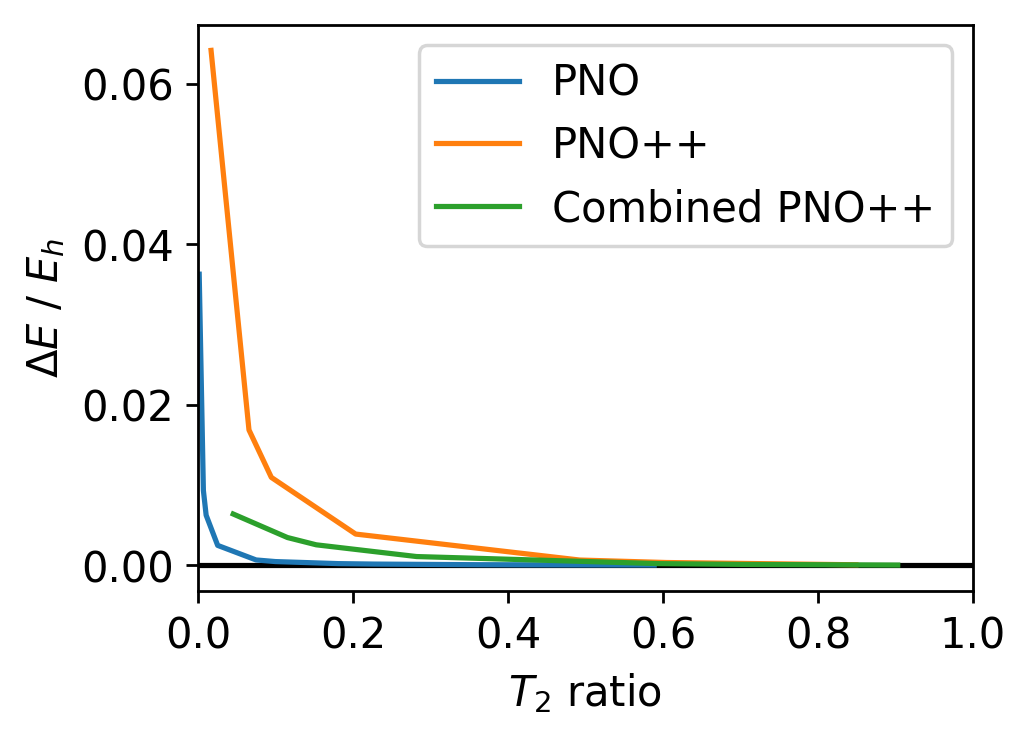}
    }
    \subfloat[]{
        \includegraphics[width=0.32\textwidth]{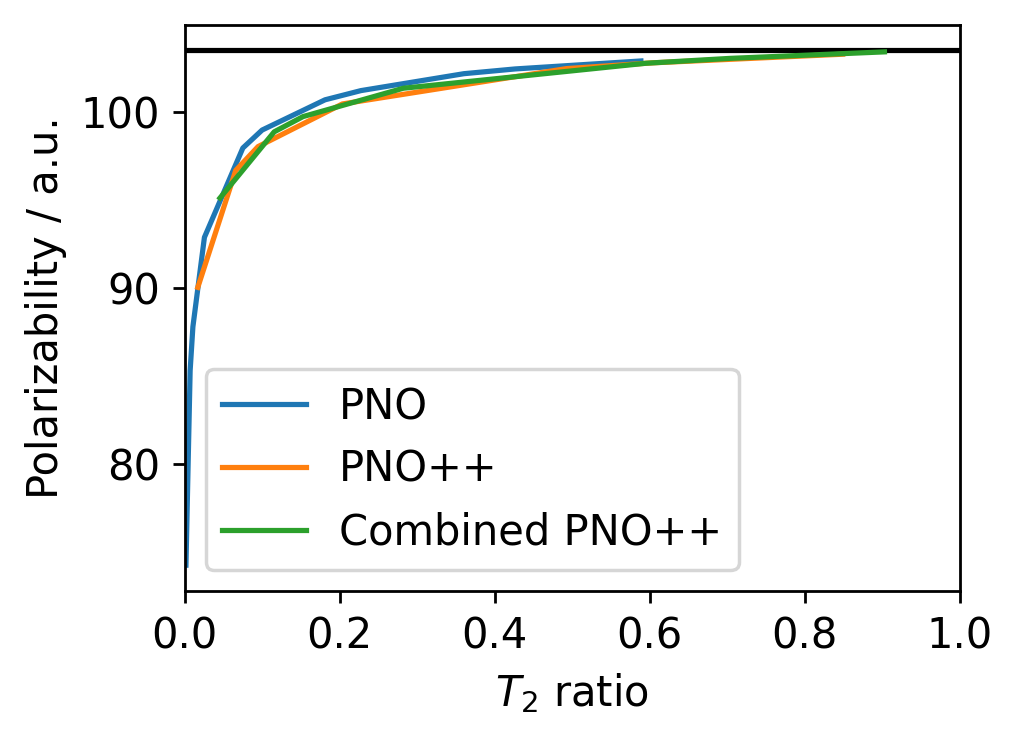}
    }
    \subfloat[]{
        \includegraphics[width=0.32\textwidth]{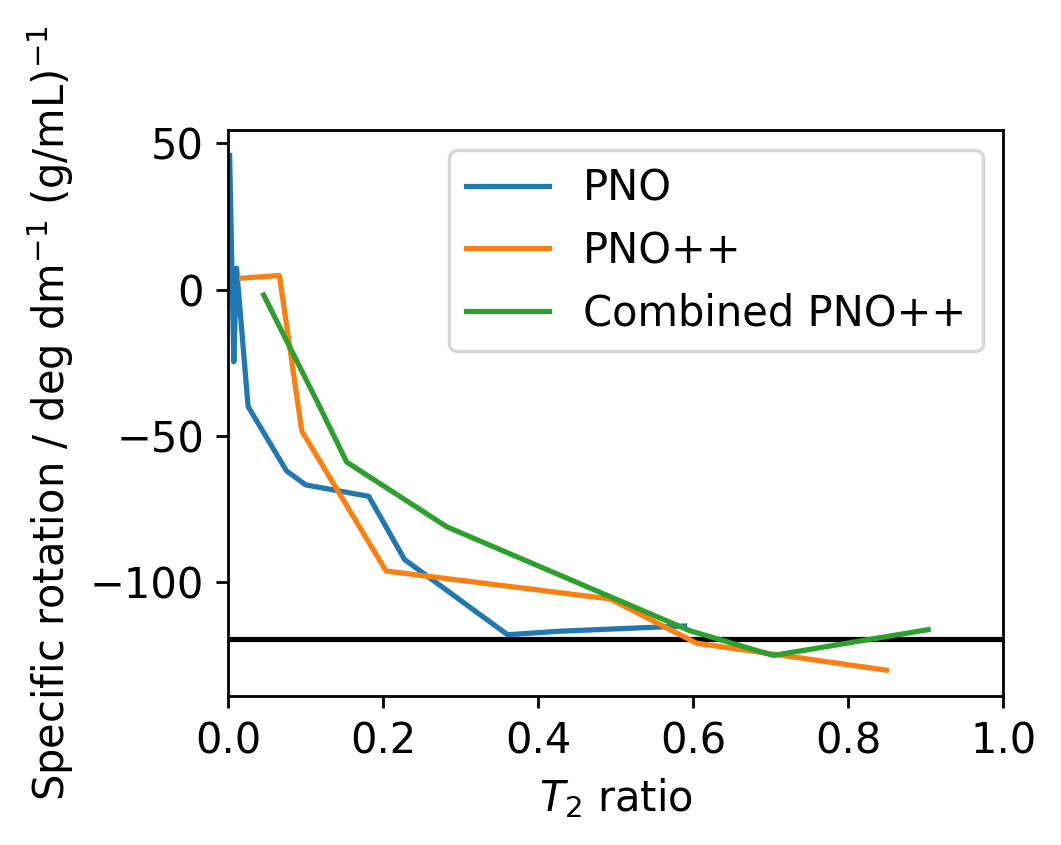}
    }
    \caption{ (a) Truncation errors in CCSD correlation energy in Hartree, (b) CCSD dynamic polarizabilities at 589 nm in a.u., and (c) CCSD specific rotations in deg dm$^{-1}$ (g/mL)$^{-1}$ at 589 nm for (\textit{S})-1-phenylethanol, computed using the aug-cc-pVDZ basis set.}
    \label{phenylethanol}
\end{figure*}

Figure \ref{phenylethanol} plots the truncation errors in correlation energy as well as the
dynamic polarizabilities and specific rotations for all three methods for 
(\textit{S})-1-phenylethanol, a challenging system for reduced scaling methods exploiting spatial
locality due to the presence of the aromatic ring substituent.  The PNO method, along with
exhibiting lower errors in correlation energy relative to the PNO++ and combined methods, also shows
similar errors for the polarizability (within 2 a.u.) to the PNO++ and combined methods. 
The convergence behavior of the specific rotation is better for the PNO method for this system
than the PNO++ method, with a point at a $T_2$ ratio of 0.36 falling within 3 deg dm$^{-1}$
(g/mL)$^{-1}$ of the reference value.  In comparison, the PNO++ and combined methods require
$T_2$ ratios of between 0.6-0.7 to achieve similarly close values. The system presents a
challenge to the local correlation methods tested here, with very large portions of the space
required in order to minimize the error in specific rotation value.

\subsection{Product Densities}

\subsubsection{Correlation Energies}

\begin{figure*}[htbp]
    \centering
    \subfloat[]{
        \includegraphics[width=0.45\textwidth]{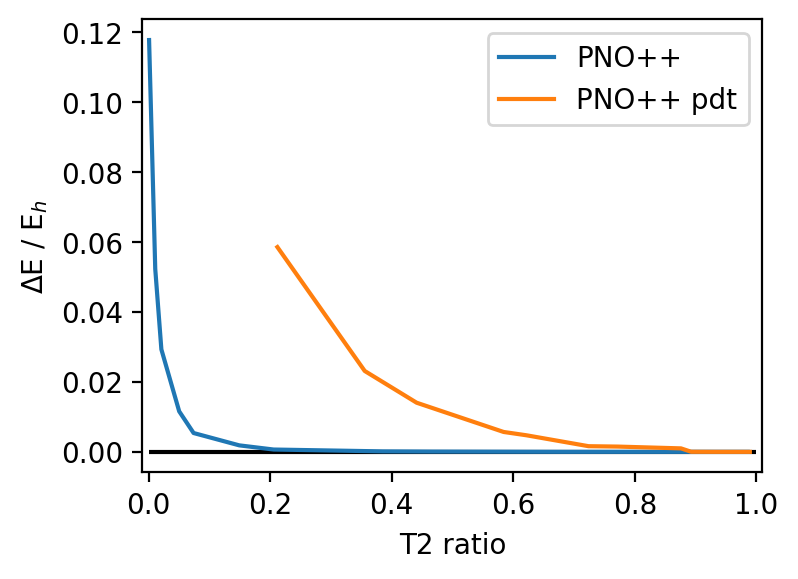}
    }
    \subfloat[]{
        \includegraphics[width=0.45\textwidth]{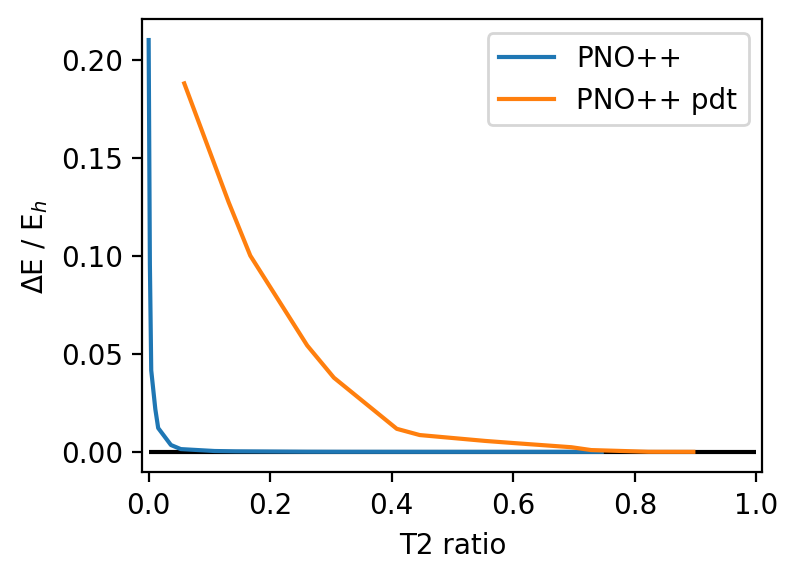}
    }
 \\
    \subfloat[]{
        \includegraphics[width=0.45\textwidth]{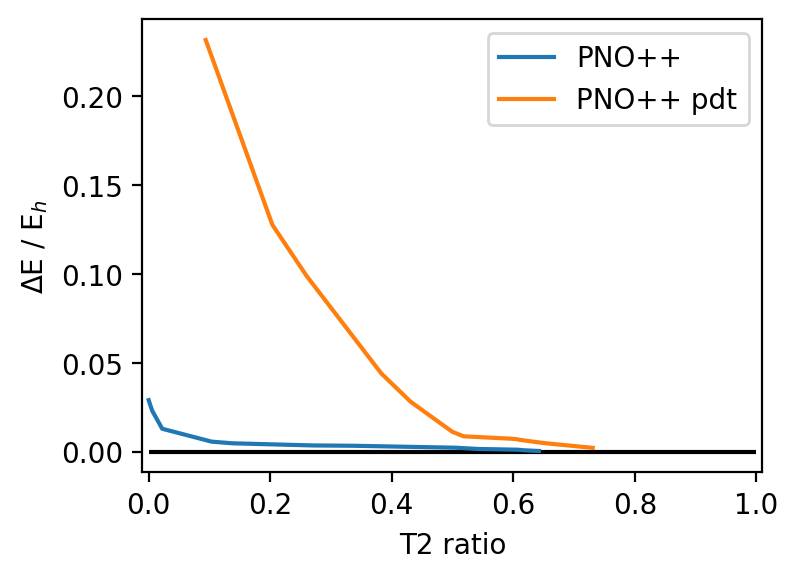}
    }
    \subfloat[]{
        \includegraphics[width=0.45\textwidth]{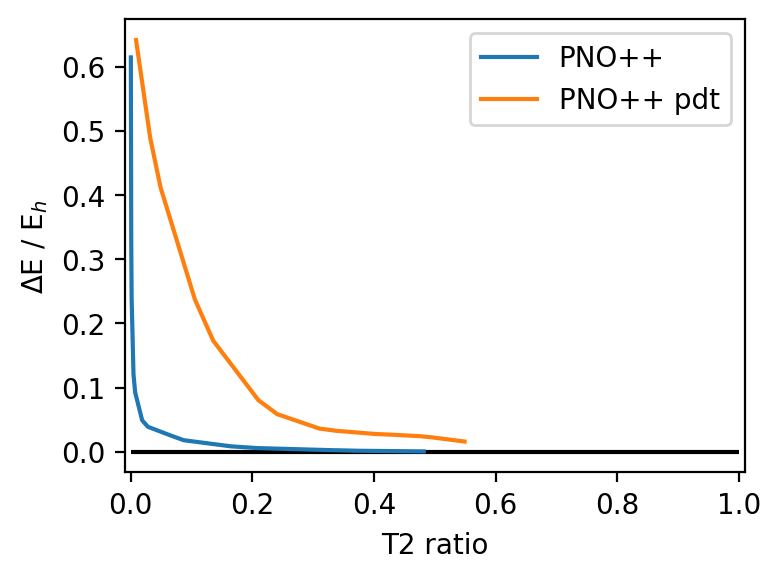}
    }
    \caption{Truncation errors in CCSD correlation energy in Hartree for (a) (H$_2$)$_4$, (b) (H$_2$)$_7$, (c) H$_2$O$_2$ and (d) DMA systems, computed using the aug-cc-pVDZ basis set for the PNO++ methods with the regular perturbed (PNO++, blue) and product (PNO++ pdt, orange) densities.}
    \label{pdt_en}
\end{figure*}

Figure \ref{pdt_en} contrasts the PNO++ method explored in Ref~\citenum{DCunha2021} with the a
similar method employing the product density described in Section \ref{product_densities} for
four small test systems: two hydrogen molecule helices, H$_2$O$_2$ and 1,3-dimethylallene (DMA)
(see Ref\citenum{DCunha2021} for geometry details).  At the same $T_2$ ratio, the truncation
error in the correlation energy is higher for the product-based density than for the regular
PNO++ density using only the electric dipole moment operator.  The product-based density requires
$T_2$ ratios greater than 0.5 in order to maintain reasonable accuracy in the correlation energy.
Since the PNO++ method itself shows higher truncation errors than the PNO method for correlation
energies at small $T_2$ ratios, this indicates that the product-based density is not a good
method for obtaining correlation energies at truncated system sizes.

\subsubsection{Optical Rotation}

\begin{figure*}[htbp]
    \centering
    \subfloat[]{
        \includegraphics[width=0.45\textwidth]{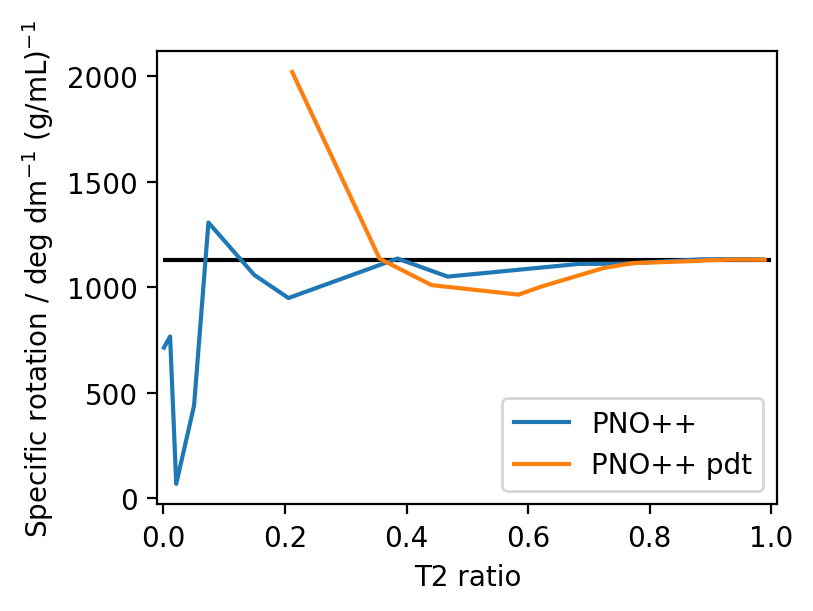}
    }
    \subfloat[]{
        \includegraphics[width=0.45\textwidth]{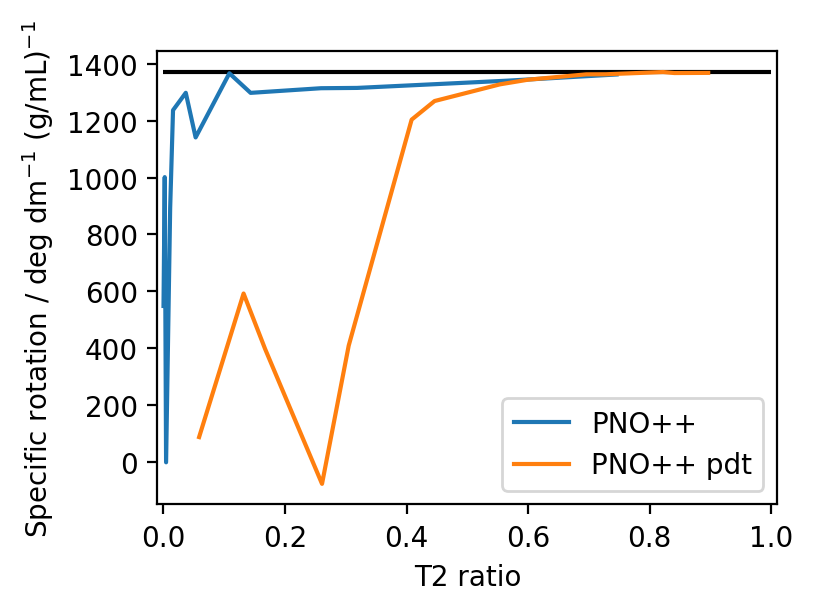}
    }
 \\
    \subfloat[]{
        \includegraphics[width=0.45\textwidth]{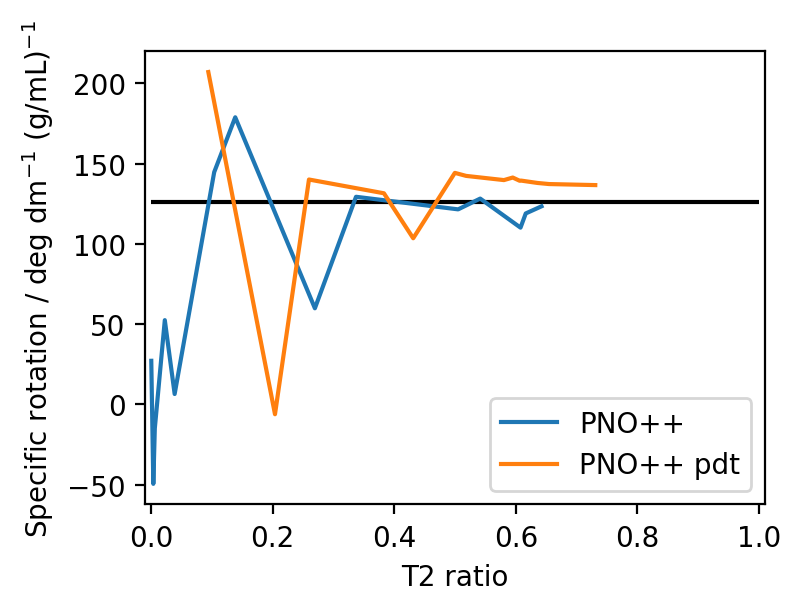}
    }
    \subfloat[]{
        \includegraphics[width=0.45\textwidth]{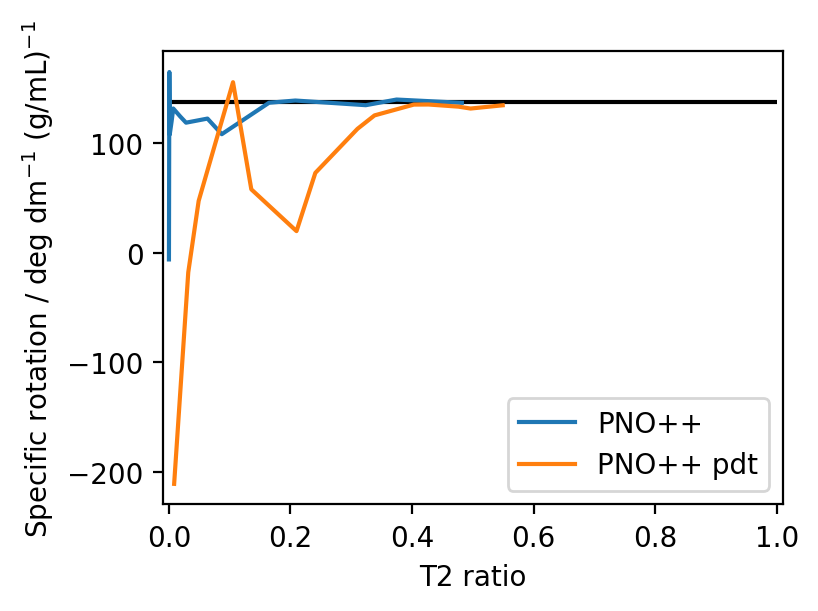}
    }
    \caption{CCSD specific rotations in deg dm$^{-1}$ (g/mL)$^{-1}$ at 589 nm for (a) (H$_2$)$_4$, (b) (H$_2$)$_7$, (c) H$_2$O$_2$ and (d) DMA systems, computed using the aug-cc-pVDZ basis set for the PNO++ methods with the regular perturbed (PNO++, blue) and product (PNO++ pdt, orange) densities.}
    \label{pdt_mvg}
\end{figure*}

Figure \ref{pdt_mvg} shows optical rotations computed using the PNO++ method with regular and
product densities. While convergence is seen for both methods, it can be seen from the figures
that the product-based density offers no improvement to the convergence behavior shown by the
PNO++ method for the four test systems.  At $T_2$ ratios below 0.25, we see large magnitude
errors in the rotation value for the product density space, with the specific rotation for DMA at
very large truncation being the wrong sign. Thus the space created by the product density is not
more optimized for the mixed property of rotation than the density containing only a single
external perturbation operator.  One possible reason for the poor behavior of the product
density versus the density using only the electric dipole moment operator may be the slower
convergence of the magnetic dipole operator with respect to basis set completeness (cf.\ the
Supporting Information of Ref~\citenum{DCunha2021}).

\subsection{Weak Pairs}

Table \ref{falkane_wp_e} contains $T_2$ ratios and errors in correlation energies for the pair
energy criterion $\epsilon$ as well as the perturbation-including criterion $\bar{\mu}$ at a
given weak pair threshold $T_{cutPairs}$ for the 1-fluoropropane, -pentane, and -heptane systems.
As the system size increases, we see the $T_2$ ratio becoming uniformly smaller at a given
threshold for the $\epsilon$ criterion indicating that fewer pairs have contributions to the
energy larger in magnitude than the chosen cutoff. This is also true for the $\bar{\mu}$
criterion and aligns with the discussion in Refs~\citenum{Russ2008} and
\citenum{McAlexander2012} that on increasing molecular sizes the computational efficiency of
local correlation methods improves.

For both of the considered criteria, errors in the correlation energy require $T_2$ ratios above
0.83 in order to remain below chemical accuracy or 1.6 m$E_h$, thus keeping a significant portion
of the space.  However, the errors in correlation energy seen using the $\bar{\mu}$ criterion are
larger at a given $T_2$ ratio than the errors using the $\epsilon$ criterion.  As an example, for
the very similar $T_2$ ratios of 0.805 and 0.795 for 1-fluoropentane, the error in correlation
energy was less than 1 m$E_h$ for the pair energy criterion, while it was above 13 m$E_h$ for the
perturbation-including criterion.  The pair energy criterion thus provides a better way to
identify weak pairs specifically for maintaining accuracy in the correlation energy.
\begin{table}[hbtp]
\centering
\begin{tabularx}{0.8\textwidth}{c|cXXcXX}
\hline
 \\
    \multicolumn{1}{c|}{Molecule}  & \multicolumn{1}{c}{Threshold} & \multicolumn{2}{c}{$\epsilon_{ij}$} && \multicolumn{2}{c}{$\bar{\mu}_{ij}$} \\ \cline{3-4}\cline{6-7}
     & & $T_2$ \mbox{ratio} & $\Delta E$ && $T_2$ \mbox{ratio} & $\Delta E$ \\ \hline \hline
    \multirow{4}{*}{1-fluoropropane} 
    & $10^{-3}$       & 0.211     & 51.1838       && 0.457       & 136.677 \\
    & $10^{-4}$       & 0.612     & 2.75454       && 0.827       & 2.64050   \\
    & $10^{-5}$       & 0.709     & 1.57325       && 0.958       & 0.22850    \\ 
    & $10^{-6}$       & 0.882     & 0.11981       && 1.000       & 0.00000  \\ \hline
    \multirow{4}{*}{1-fluoropentane} 
    & $10^{-3}$       & 0.146     & 83.3302       && 0.413       & 70.5225  \\
    & $10^{-4}$       & 0.456     & 6.20241       && 0.795       & 13.6891   \\
    & $10^{-5}$       & 0.568     & 3.20850       && 0.955       & 0.05368    \\ 
    & $10^{-6}$       & 0.805     & 0.15424       && 0.987       & 0.00912  \\ \hline
    \multirow{4}{*}{1-fluoroheptane} 
    & $10^{-3}$       & 0.111     & 112.815       && 0.328       & 87.2679  \\
    & $10^{-4}$       & 0.361     & 9.77172       && 0.722       & 16.5131   \\
    & $10^{-5}$       & 0.469     & 4.85166       && 0.935       & 2.22753    \\ 
    & $10^{-6}$       & 0.726     & 0.18945       && 0.991       & 0.01013 \\ \hline
\end{tabularx}
\caption{$T_2$ ratios and errors in correlation energy (m$E_h$) computed at the CCSD level for the three 1-fluoroalkane systems as a function of the $T_{cutPairs}$ threshold, using the aug-cc-pVDZ basis set.}
\label{falkane_wp_e}
\end{table}

Table \ref{falkane_wp_or} presents polarizabilities and specific rotations computed using the
modified velocity gauge as a function of the threshold for both criteria for the 1-fluoropropane,
-pentane, and -heptane systems.  For all three fluoroalkanes and all values of the threshold the
polarizability value remains within 1 a.u. of the reference, even at the smallest $T_2$ ratios.
Thus the polarizability is not strongly affected by the neglect of weak pairs using either
criterion.  The specific rotation, on the other hand, is affected by the truncation and is seen
to require a $T_2$ ratio between 0.5-0.7 for the $\epsilon$ criterion and 0.4-0.8 for the
$\bar{\mu}$ criterion to remain within 10\% of the reference CCSD value.

\begin{table}[hbtp]
\centering
\begin{tabularx}{\textwidth}{c|ccXXccXX}
\hline
  \\
    \multicolumn{1}{c|}{Molecule}  & \multicolumn{1}{c}{Threshold} & \multicolumn{3}{c}{$\epsilon_{ij}$} && \multicolumn{3}{c}{$\bar{\mu}_{ij}$} \\ \cline{3-5}\cline{7-9}
     & & $T_2$ \mbox{ratio} & $\alpha$ & MVG && $T_2$ \mbox{ratio} & $\alpha$ & MVG \\ \hline \hline
    \multirow{5}{*}{1-fluoropropane} 
    & $10^{-3}$       & 0.211     & 42.02       & \mbox{-83.06} && 0.457       & 42.69      & \mbox{-58.10} \\
    & $10^{-4}$       & 0.612     & 41.69       & \mbox{-77.05} && 0.827       &41.67       & \mbox{-60.30}  \\
    & $10^{-5}$       & 0.709     & 41.66       & \mbox{-64.56} && 0.958       &41.66       & \mbox{-57.66}   \\ 
    & $10^{-6}$       & 0.882     & 41.65       & \mbox{-63.12} && 1.000       &41.65       & \mbox{-62.89} \\ \hline
    \multirow{5}{*}{1-fluoropentane} 
    & $10^{-3}$       & 0.146     & 67.26       & \mbox{-77.99}  && 0.413       & 67.61      & \mbox{-71.85}  \\
    & $10^{-4}$       & 0.456     & 66.64       & \mbox{-65.14}  && 0.795       &66.95       & \mbox{-57.39} \\
    & $10^{-5}$       & 0.568     & 66.77       & \mbox{-57.89}  && 0.955       &66.77       & \mbox{-54.78}  \\ 
    & $10^{-6}$       & 0.805     & 66.77       & \mbox{-54.96}  && 0.987       &66.77       & \mbox{-54.11}\\ \hline
    \multirow{5}{*}{1-fluoroheptane} 
    & $10^{-3}$       & 0.111     & 92.73       & \mbox{-62.17}  && 0.328       & 92.99       & \mbox{-50.76}  \\
    & $10^{-4}$       & 0.361     & 91.89       & \mbox{-53.42}  && 0.722       & 92.47       & \mbox{-43.19} \\
    & $10^{-5}$       & 0.469     & 92.21       & \mbox{-47.57}  && 0.935       & 92.30       & \mbox{-42.51}  \\ 
    & $10^{-6}$       & 0.726     & 92.30       & \mbox{-44.82}  && 0.991       & 92.30       & \mbox{-44.21} \\ \hline
\end{tabularx}
    \caption{$T_2$ ratios, dynamic polarizabilities (a.u.), and specific rotations (deg dm$^{-1}$ (g/mL)$^{-1}$) at 589 nm computed at the CCSD level for the three 1-fluoroalkane systems as a function of the $T_{cutPairs}$ threshold, using the aug-cc-pVDZ basis set. $\alpha_{\textrm{Ref}}$ 1-fluoropropane: 41.65, 1-fluoropentane: 66.77, 1-fluoroheptane: 92.30. MVG$_{\textrm{Ref}}$ 1-fluoropropane: -63.03, 1-fluoropentane: -53.72, 1-fluoroheptane: -43.75.}
\label{falkane_wp_or}
\end{table}

\section{Conclusions}

In this work, we have compared the performance of the PNO++ and combined PNO++ methods to the conventional PNO method for larger molecules
than considered in previous work using a new implementation within the Psi4 package.  We have also tested a new product-based density in
the creation of the PNO++ space, as well as the accuracy limits of both an energy- and perturbation-based weak-pair threshold.

For the series of 1-fluoroalkane systems tested, we see better convergence behavior and lower truncation errors for the PNO++ and combined
PNO++ methods than the PNO for the linear response properties tested.  While the truncation errors in correlation energy were large for the
PNO++ method, we recover similar accuracy to the PNO method by incorporating a number of the original pair natural orbitals in the combined
PNO++ space.  The bicyclic $\alpha$- and $\beta$-pinene molecules appear to be more difficult test cases than the linear alkane chains for
all three methods tested, requiring large amounts of the virtual space to be retained in order to obtain accuracy in the dynamic
polarizability, and showing large errors in the specific rotation value even at small truncations.  A surprising result is the very similar
convergence behavior of the polarizability for all three methods, with all three having approximately the same error at a given truncation.
Finally, truncation errors in both the correlation energy and the specific rotation are smaller for the PNO method than the PNO++ or combined
methods for the system of (\textit{S})-1-phenylethanol.

The use of a product perturbed density, in which mixed perturbations are incorporated simultaneously, was tested for its ability to produce a
more compact space for specific rotations, having combined two perturbed densities via a Hadamard product. However, the results suggest that
the density using a single electric-field perturbation is more effective than the product density and that further testing --- particularly
for the basis-set completeness of the magnetic dipole operator --- is required in order to understand the impact of using a mixed density in
order to create the PNO++ space.

The two weak pair criteria examined here underscore several observations from the larger benchmark calculations with the PNO, PNO++, and
combined PNO++ methods.  First, $T_2$ ratios are larger at a given threshold for the $\bar{\mu}$ than for the $\epsilon$ criterion,
suggesting that this criterion naturally contains larger magnitude contributions than the pair energy criterion.  Second, the $\epsilon$
criterion provides a better way to identify weak pairs specifically for maintaining accuracy in the correlation energy, while the
polarizability is not affected by aggressive truncation, and thus either criterion could be used to truncate the space.  For the specific
rotation, the ability to truncate at a lower threshold for a given criterion is system-dependent, with the linear fluoroalkane chains
requiring slightly less of the space with the $\bar{\mu}$ than with the $\epsilon$ criterion.  This mirrors observations of the truncation
error with the PNO method versus the PNO++ method for the same systems. 

Through this study we have studied the effect of truncation on CCSD correlation energies, dynamic polarizabilities, and specific rotations
using three local correlation methods, the PNO, PNO++ and combined PNO++ methods.  While the results for the linear alkane chains are
encouraging in their better convergence behavior, it is clear that the PNO++ method is not yet a panacea for the local-correlation
computation of linear-response properties.  The reasons for the inconsistent behavior of all of the PNO-based methods --- from conventional
PNO, to the various flavors of the PNO++ method considered here -- require further investigation.

\section{Acknowledgements}

This research was supported by the U.S. National Science Foundation (grant
CHE-1900420). The authors are grateful to Advanced Research Computing at
Virginia Tech for providing computational resources and technical support
that have contributed to the results reported within the paper.

\bibliographystyle{tfo}
\bibliography{local_correlation,thesis_refs,refs}

\end{document}